\documentclass[sigconf]{acmart}
\usepackage{float}                  % 图片浮动位置
\usepackage{subfigure}                 % 子图包，不要与{subfigure}混用，{subfig}较新
\usepackage{overpic}                % 与图片排版相关
\usepackage{graphicx}
\usepackage{multirow}
\usepackage{marvosym}
\usepackage[labeled]{multibib}
\newcites{A}{Appendix References}

\acmSubmissionID{3609}
\copyrightyear{2023}
\acmYear{2023}
\setcopyright{rightsretained}
\acmConference[MM '23]{Proceedings of the 31st ACM International Conference on Multimedia}{October 29-November 3, 2023}{Ottawa, ON, Canada}
\acmBooktitle{Proceedings of the 31st ACM International Conference on Multimedia (MM '23), October 29-November 3, 2023, Ottawa, ON, Canada}
\acmDOI{10.1145/3581783.3612503}
\acmISBN{979-8-4007-0108-5/23/10}

\begin{document}

\title{UnifiedGesture: A Unified Gesture Synthesis Model for Multiple Skeletons}


% \author{Anonymous authors\\Paper ID 123456}

\author{Sicheng Yang}
\authornote{Equal contribution}
\affiliation{Shenzhen International Graduate School, Tsinghua University
\city{Shenzhen}
\country{China}}
\email{yangsc21@mails.tsinghua.edu.cn}
\orcid{0000-0002-0928-034X}

\author{Zilin Wang}
\authornotemark[1]
\affiliation{Shenzhen International Graduate School, Tsinghua University
\city{Shenzhen}
\country{China}}
\email{wangzl21@mails.tsinghua.edu.cn}
\orcid{0009-0003-6062-3015}

\author{Zhiyong Wu}     % \textsuperscript{\Letter}
\authornote{Corresponding author}
\affiliation{Shenzhen International Graduate School, Tsinghua University
\city{Shenzhen}
\country{China}}
\affiliation{
The Chinese University of Hong Kong
\city{Hong Kong SAR}
\country{China}}
\email{zywu@sz.tsinghua.edu.cn}
\orcid{0000-0001-8533-0524}

% \author{Minglei Li\textsuperscript{$\dagger$}}
\author{Minglei Li}
\authornotemark[2]
\affiliation{Huawei Cloud Computing Technologies Co., Ltd
\\\city{Shenzhen}
\country{China}}
\email{liminglei29@huawei.com}
\orcid{0000-0002-1427-3507}

\author{Zhensong Zhang}
\affiliation{Huawei Noah’s Ark Lab
\\\city{Shenzhen}
\country{China}}
\email{zhangzhensong@huawei.com}
\orcid{0009-0001-7911-7564}

\author{Qiaochu Huang}
\affiliation{Shenzhen International Graduate School, Tsinghua University
\city{Shenzhen}
\country{China}}
\email{hqc22@mails.tsinghua.edu.cn}
\orcid{0009-0004-8113-6459}

\author{Lei Hao}
\affiliation{Huawei Noah’s Ark Lab
\\\city{Shenzhen}
\country{China}}
\email{haolei5@huawei.com}
\orcid{0009-0009-6977-119X}

\author{Songcen Xu}
\affiliation{Huawei Noah’s Ark Lab
\\\city{Shenzhen}
\country{China}}
\email{xusongcen@huawei.com}
\orcid{0000-0002-0022-0906}

\author{Xiaofei Wu}
\affiliation{Huawei Noah’s Ark Lab
\\\city{Shenzhen}
\country{China}}
\email{wuxiaofei2@huawei.com}
\orcid{0009-0007-0143-1485}

\author{Changpeng Yang}
\affiliation{Huawei Cloud Computing Technologies Co., Ltd
\\\city{Shenzhen}
\country{China}}
\email{yangchangpeng@huawei.com}
\orcid{0000-0002-7043-6657}

\author{Zonghong Dai}
\affiliation{Huawei Cloud Computing Technologies Co., Ltd
\\\city{Shenzhen}
\country{China}}
\email{daizonghong@huawei.com}
\orcid{0009-0006-7723-4130}

\def\shortauthors{Sicheng Yang, et al.}

%% DeepGesture: Audio-Driven Co-Speech Gesture Generation with Diffusion Models 
% DeepGesture: Diffusion model-based and reinforcement learning-guided speech-driven gesture generation training on mutiple datasets

\begin{abstract}
  The automatic co-speech gesture generation draws much attention in computer animation.
  Previous works designed network structures on individual datasets, which resulted in a lack of data volume and generalizability across different motion capture standards.
  In addition, it is a challenging task due to the weak correlation between speech and gestures.
  To address these problems, we present UnifiedGesture, a novel diffusion model-based speech-driven gesture synthesis  approach, trained on multiple gesture datasets with different skeletons.        % and reinforcement learning-guided 
  Specifically, we first present a retargeting network to learn latent homeomorphic graphs for different motion capture standards, unifying the representations of various gestures while extending the dataset. 
  We then capture the correlation between speech and gestures based on a diffusion model architecture using cross-local attention and self-attention to generate better speech-matched and realistic gestures.
  %Moreover, we introduce a gesture VQ-VAE to learn a codebook to summarize meaningful gesture units, with each code representing a unique gesture. 
  To further align speech and gesture and increase diversity, we incorporate reinforcement learning on the discrete gesture units with a learned reward function.
  Extensive experiments show that UnifiedGesture outperforms recent approaches on speech-driven gesture generation in terms of CCA, FGD, and human-likeness.
  All code, pre-trained models, databases, and demos are available to the public at \url{https://github.com/YoungSeng/UnifiedGesture}.

% Automatic co-speech gesture generation is an important topic in computer animation, but previous work has suffered from a lack of data volume and generalizability across different motion capture standards, as well as the challenge of the weak correlation between speech and gestures.

    % In this paper, we propose a novel approach to address the lack of exploration capability in current music-conditioned 3D dance generation models. Our approach involves training a reward model from automatically-ranked dance demonstrations, which is then used to guide the reinforcement learning process that encourages the dance agent to explore and generate more diverse dance movement sequences. The soundness of our reward model is both theoretically and experimentally validated. Additionally, our proposed dance generation framework, E3D2, achieves state-of-the-art performance on the AIST++ dataset. 
  
\end{abstract}




\maketitle

\section{Introduction} % 1.5-2 pages    

\begin{figure}[t]
  \centering
  \includegraphics[width=\linewidth]{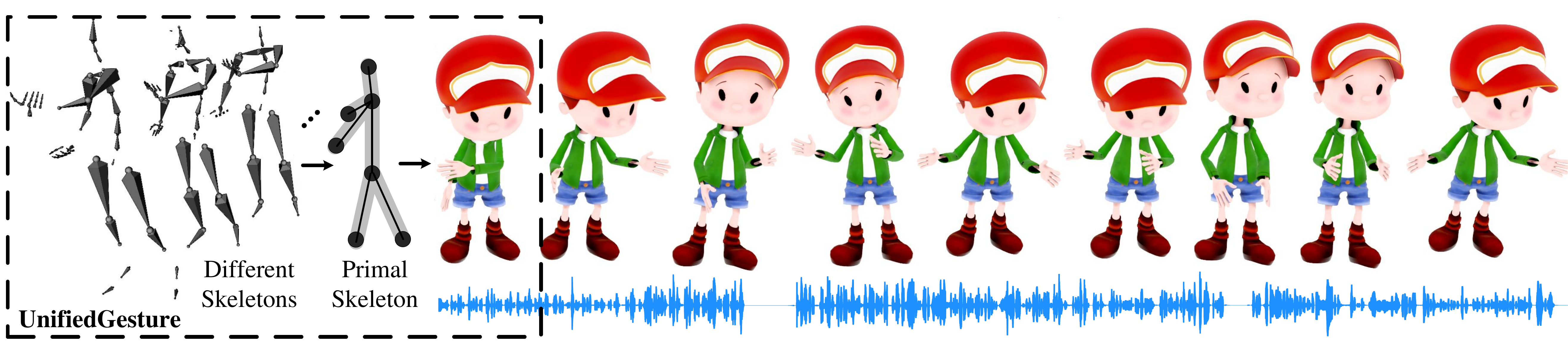}
  \caption{Gesture examples generated by our proposed method.
  Different skeletons are unified to the primal skeleton.
  The speech-driven primal skeleton generate gestures for the specified skeleton.
  The character used in the paper is publicly available.
  }
  \label{fig:teaser}
\end{figure}

Nonverbal behaviors, including gestures, play key roles in conveying messages in human communication \cite{DBLP:conf/iui/KucherenkoJYWH21}.
The automatic co-speech gesture generation is considered an enabling technology to create realistic 3D avatars in films, games, virtual social spaces, and for interaction with social robots \cite{DBLP:journals/corr/abs-2301-05339}. 
In the era of deep learning, existing data-driven gesture generation methods usually rely on a large dataset. Studies have shown that a larger amount of data can improve the generalization of the model and enhance its performance \cite{DBLP:journals/corr/abs-2209-03143, DBLP:journals/corr/abs-2303-08774}. 

Thanks to the development of human pose estimation~\cite{li2022cliff}, it's easy to extract 3D human poses from tremendous 2D gesture data on the web, e.g., TED \cite{DBLP:conf/icra/YoonKJLKL19} and PATS \cite{DBLP:conf/emnlp/AhujaLIM20}, some works \cite{DBLP:conf/icra/YoonKJLKL19, DBLP:journals/tog/YoonCLJLKL20, DBLP:conf/cvpr/LiangFZHP022} are based on 2D gesture datasets.      %  DBLP:conf/cvpr/Ahuja0M22
While large in quantity, 3D poses extracted from 2D datasets are poor in quality and difficult to use, most works \cite{DBLP:conf/mig/FerstlNM19, DBLP:journals/cgf/AlexandersonHKB20, DBLP:conf/icmi/KucherenkoJWHAL20, DBLP:conf/iccv/0071KPZZ0B21, DBLP:conf/eccv/LiuZIPLZBZ22} opt for high quality 3D mocap datasets.      % DBLP:conf/cvpr/mycvpr, DBLP:journals/tog/AoGLCL22, 
%These methods for 3D co-speech gesture generation have been constructed on a single dataset, which resulted in a lack of data volume and generalizability across different motion capture standards.

There are two main challenges when utilizing 3D datasets. First, due to the expensive cost of motion capture, the typical 3D gesture datasets \cite{DBLP:journals/cgf/GhorbaniFHTC23, DBLP:conf/iva/FerstlM18, DBLP:conf/atal/KucherenkoNNKH22,DBLP:conf/eccv/LiuZIPLZBZ22} are relatively small, thus the generalization of the models trained on the individual dataset is limited, and the ability of the trained algorithms is also confined to the content of the individual dataset. For example, some datasets contain style information~\cite{DBLP:journals/cgf/GhorbaniFHTC23,DBLP:conf/eccv/LiuZIPLZBZ22}, while the others do not~\cite{DBLP:conf/iva/FerstlM18,DBLP:conf/atal/KucherenkoNNKH22}. Second, it is not straightforward to train algorithms on mixture datasets directly, since different datasets usually have different skeletons, they are captured with different mocap systems.
Most of the current solutions use software such as Blender \cite{Blender} or Maya \cite{Maya} for automatic retargeting to a unified skeleton, which requires manual specification of the bone mapping and leads to unavoidable errors \cite{DBLP:journals/tog/AbermanLLSCC20}.
The irregular connectivity and hierarchical structure of the skeleton joint motion cause difficulties in the large-scale application of multiple skeletons.

\begin{figure}[t]
  \centering
  \includegraphics[width=\linewidth]{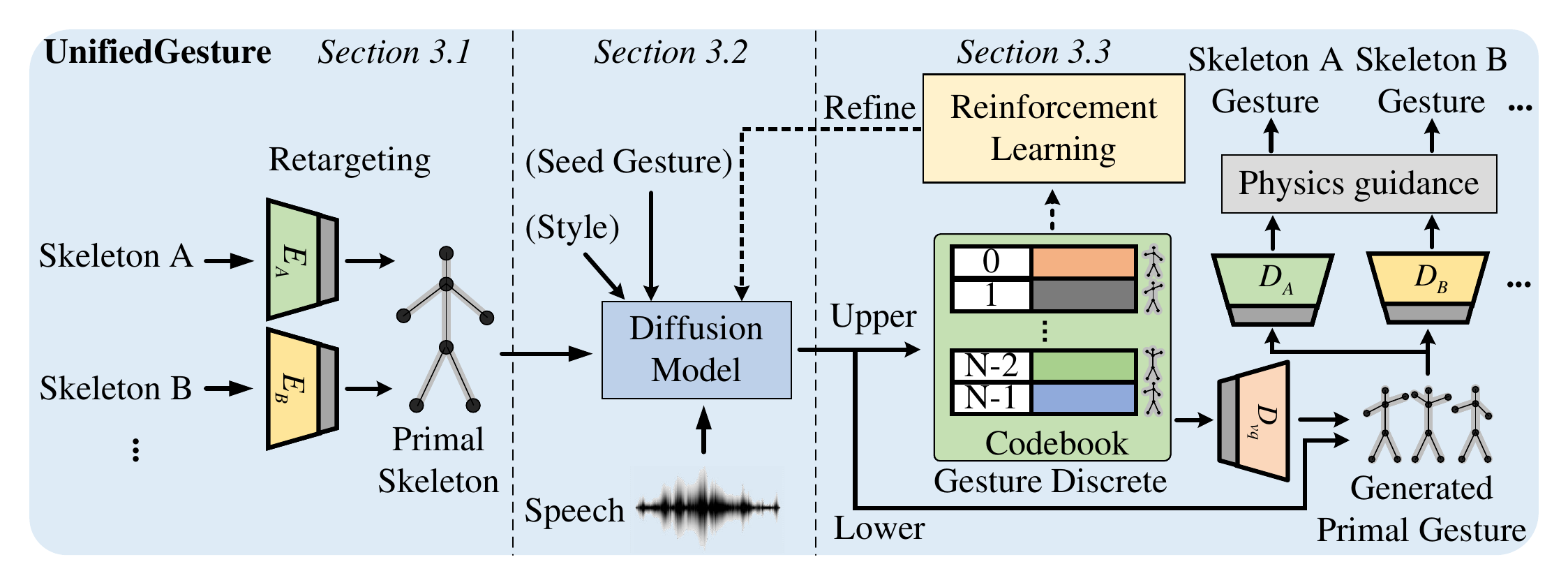}
  \caption{Gesture generation pipeline of our proposed framework. 
  We retarget the different skeletons to the primal skeleton.
  Given a speech segment (optional style, seed gesture), the output is the primal gesture after VQVAE encoding of the output of the diffusion model.
  We introduce reinforcement learning to refine the gesture generation network.
  Finally, a gesture of the skeleton is specified and generated, with physics guidance.
  }
  \Description{framework.}
    \label{framework}
\end{figure}

To tackle these challenges, we propose UnifiedGesture, a novel unified co-speech gesture synthesis model for multiple skeletons.
The overview of our method is shown in Figure \ref{framework}.
Although the number and position of the different skeleton joints are different, they all correspond to homeomorphic (topologically equivalent) graphs \cite{DBLP:conf/www/WangJSWYCY19}.
Unlike sign language or hand gestures, there is a weak correlation between speech and body gestures at a coarse-grained level \cite{DBLP:journals/corr/abs-2208-09141, DBLP:journals/corr/abs-2303-01765}. 
Specifically, we assume that the gesture details associated with speech are contained in the primal skeleton gesture.
According to this assumption, we first use a data-driven deep skeleton-aware \cite{DBLP:journals/tog/AbermanLLSCC20} framework to learn latent homeomorphic graphs for different skeletons.
The different skeletons are unified and retargeted to the primal skeleton while extending the dataset.
Then we introduce a denoising-diffusion-based speech-driven co-speech gesture generation model, using WavLM features \cite{DBLP:journals/jstsp/ChenWCWLCLKYXWZ22}, based on cross-local attention \cite{DBLP:journals/tacl/RoySVG21} and self-attention \cite{DBLP:conf/nips/VaswaniSPUJGKP17} architecture to better capture the temporal information between audio and gestures.
%Random masks are used to perform a classifier-free \cite{DBLP:journals/corr/abs-2207-12598} for interpolation and editing of conditions.
Third, unlike speaking with the face or lips, the weak correlation between speech and gesture lacks a suitable criterion for learning the model, to refine the gesture generation model, we employ inverse reinforcement learning (IRL) on discrete gesture units to train a reward model that evaluates the generated gestures and guides the diffusion model to generate high-quality and diverse gestures aligned with speech during the reinforcement learning (RL) process.
%Third, we compress primal gestures of the upper body into a space that is lower dimensional and discrete, to reduce input redundancy.
%Instead of manually indicating the gesture units \cite{DBLP:phd/de/Kipp2007}, we use a vector quantized variational autoencoder (VQVAE) \cite{DBLP:conf/nips/OordVK17} to encode and primal gestures sequences to a codebook in an unsupervised manner, using a quantization bottleneck.
%Each learned code is shown to represent a unique gesture pose.
%By reconstructing discrete gestures, the exploratory space for reinforcement learning can be reduced, while increasing controllability and interpretability.
%Further, unlike speaking with the face or lips, the weak correlation between speech and gesture lacks a suitable criterion for learning the model.
%To refine the gesture generation model, we employ inverse reinforcement learning (IRL) to train a reward model that evaluates the generated gestures and guides the diffusion model to generate high-quality and diverse gestures aligned with speech during the reinforcement learning (RL) process.
%Finally, non-physical-based motion generation systems are associated with problems such as foot sliding and root shifting \cite{DBLP:journals/corr/abs-2301-05175, DBLP:journals/cgf/GhorbaniFHTC23}.
%We design constraints based on the relationship between roots and feet \cite{DBLP:journals/corr/abs-2211-10658}, using standard inverse kinematics (IK) optimized for physics guidance.
Our code, pre-trained models, and demos will be publicly available soon.
The main contributions of our work are:
\begin{itemize}
    \item We employ a skeleton-aware retargeting network to unify the different skeletons to a common primal skeleton while extending the dataset.
    \item We present a temporally aware attention-based diffusion model on the primal skeleton for speech-driven co-speech gesture generation. 
    By virtue of the diffusion model, we can edit the style of the gestures, setting the initial gestures, and generating diverse gestures.
%    \item We use VQVAE to learn a codebook to summarize meaningful gesture units. The exploratory space for reinforcement learning can be reduced, while increasing controllability and interpretability.
    \item We introduce reinforcement learning with a learned reward function to refine the generation model and make the model explore the data. The exploratory space for reinforcement learning is reduced by learning a codebook with VQVAE to summarize meaningful gesture units.
    \item Extensive experiments show that our model can generate human-like, speech-matched, stylized, diverse, controllable, and physically plausible gestures that significantly outperform existing gesture generation methods.
\end{itemize}

\section{Related Work} % 2.5-2.75 pages

\subsection{Motion Retargeting}

Our task is to take advantage of multiple gesture datasets. 
There are two main challenges. 
First, different datasets have the same motion capture standards (e.g., Trinity \cite{DBLP:conf/iva/FerstlM18} and BEAT \cite{DBLP:conf/eccv/LiuZIPLZBZ22} both using Vicon’s suits); second, different datasets have different motion capture standards (e.g., ZEGGS \cite{DBLP:journals/cgf/GhorbaniFHTC23} and Talking With Hands \cite{DBLP:conf/iccv/LeeDMSSS19}).
For the first case, we can select the body joints common to both datasets and unify the different skeletons normalized \cite{DBLP:journals/tog/YoonCLJLKL20} by height or arm span, etc.
The latter case is more challenging. 
\citet{DBLP:journals/corr/abs-2212-02837} try to map multiple datasets to a defined skeleton, but it still partially relies on handcrafting and the results are still limited to a specific motion skeleton.
Some work \cite{DBLP:journals/jvca/KimJK20, DBLP:conf/iva/KucherenkoHHKK19} try to retarget the motion of different skeletons by VAE, using standard convolution and pooling.
However, unlike images or videos, different skeletons exhibit irregular connectivity.
\citet{DBLP:conf/cvpr/VillegasYCL18} propose a neural network for motion retargeting that adapts input motion to target characters, achieving state-of-the-art results and using cycle consistency for unsupervised learning.
\citet{DBLP:conf/bmvc/LimCC19} propose a pose-movement network for motion retargeting using a normalizing process and novel loss function.
\citet{DBLP:journals/cgf/KimPKH20} present an unsupervised motion retargeting model using temporal dilated convolutions that generates realistic and stable trajectories for humanoid characters.
\citet{DBLP:conf/iccv/VillegasCHYS21} propose a motion retargeting method that preserves self-contacts and prevents interpenetration, using a recurrent network. 
\citet{DBLP:journals/cg/LiWJZZ22} propose an iterative motion retargeting method using an iterative motion retargeting network for unsupervised motion retargeting.
Inspired by \cite{DBLP:journals/tog/AbermanLLSCC20}, we use a deep skeleton-aware framework for data-driven motion retargeting between skeletons.

\subsection{Gesture Generation}

\subsubsection{End-to-end Co-speech Gesture Generation}
Gesture generation is a complex task that requires understanding speech, gestures, and their relationships.
The present data-driven studies mainly consider four modalities: text \cite{DBLP:conf/iva/AlexandersonSHK20, DBLP:conf/icra/YoonKJLKL19, DBLP:conf/icmi/WangAGBHS21}, audio \cite{DBLP:conf/cvpr/GinosarBKCOM19, DBLP:conf/iva/HabibieXMLSPET21, DBLP:conf/iccv/QianTZ0G21}, gesture motion \cite{DBLP:journals/tog/YoonCLJLKL20, DBLP:conf/icmi/LuF22, DBLP:conf/ijcai/myijcai}, and speaker identity \cite{DBLP:conf/eccv/AhujaLNM20, DBLP:journals/cgf/AlexandersonHKB20, DBLP:conf/cvpr/LiuWZXQLZWDZ22}.
There are some works to extend the scale of the dataset.
\citet{DBLP:conf/eccv/LiuZIPLZBZ22} present a large motion capture dataset for studying the correlation of conversational gestures with facial expressions, emotions, and semantics.
% and propose cascaded motion network to synthesize gesture.
\citet{DBLP:conf/atal/KucherenkoNNKH22} provide further annotation of the specific properties and details of the gestures in the dataset.
\citet{DBLP:journals/cgf/GhorbaniFHTC23} propose a dataset containing motion styles compared to the previous dataset containing speech styles.
However, these methods are currently only tried on a single dataset, which resulted in a lack of data volume and generalizability across different motion capture standards. 

Some works \cite{DBLP:conf/siggraph/HabibieESANNT22,DBLP:conf/icmi/ZhouBC22} use motion-matching methods to generate co-speech gestures.
\citet{DBLP:conf/cvpr/mycvpr} try to transform the original gesture motion space to the deep latent phase space \cite{DBLP:journals/tog/StarkeMK22}, but it is still based on the traditional convolutional network, ignoring the hierarchy and connectivity between the skeletons.
Besides, this approach requires careful design of the database, which is directly related to the performance of the generated gestures.
The length of matching needs to be balanced between quality and diversity. 
Furthermore, the approach also requires the complex and time-consuming manual design of the matching rules.

\subsubsection{Diffusion Models for Motion Generation.}
Diffusion models \cite{DBLP:conf/nips/HoJA20} excel at modeling complicated data distribution and generating vivid motion sequences.
Many works \cite{DBLP:journals/corr/abs-2209-00349, DBLP:journals/corr/abs-2210-12315, DBLP:journals/corr/abs-2209-14916} integrate diffusion-based generative models into the motion domain.       % DBLP:journals/corr/abs-2303-01418, DBLP:journals/corr/abs-2302-05905, DBLP:journals/corr/abs-2212-02500
There are some works \cite{DBLP:conf/mmm/ZhangJGL23, DBLP:journals/corr/abs-2211-09707, DBLP:journals/corr/abs-2303-09119} that introduce diffusion models in gesture generation to demonstrate the potential of diffusion models in solving cross-modal, time-series relations problems.     % DBLP:conf/ijcai/myijcai
In our work, we use a well-designed attention architecture in the diffusion model to make the generated gestures match better with the speech.

\subsubsection{Quantization-based Pose Representation.}
Kipp has represented gestures as predefined unit gestures \cite{DBLP:phd/de/Kipp2007}.
% \citet{DBLP:conf/nips/OordVK17} propose Vector VQ-VAE to generate discrete representations.
% \citet{DBLP:journals/tog/HongZPCYL22} use a pose codebook created by clustering to generate diverse poses.
\citet{DBLP:conf/eccv/LucasBWR22} propose to train a GPT-like model for next-index prediction.
\citet{DBLP:conf/cvpr/SiyaoYGLW0L022} propose to pose VQ-VAE \cite{DBLP:conf/nips/OordVK17} to encode and summarize dancing units.
In terms of gesture generation, there are several works \cite{DBLP:journals/tog/AoGLCL22, DBLP:journals/corr/abs-2208-09141, DBLP:journals/corr/abs-2211-16016} that apply VQVAE to encode meaningful gesture units.
Existing studies \cite{DBLP:conf/eccv/GuoZWC22, DBLP:journals/tog/HongZPCYL22, DBLP:conf/eccv/LucasBWR22} have shown that quantification helps to reduce motion freezing during motion generation and retains the details of motion well.
Unlike them, we encode the gesture units in the deep primal motion latent space.

\subsection{Reinforcement Learning}
% Reinforcement Learning (RL) has emerged as a powerful and effective approach for addressing sequential decision-making problems. RL algorithms aim to learn a policy that maximizes the cumulative reward signal through iterative interactions with the environment \cite{DBLP:books/lib/SuttonB98}. This process involves the agent taking actions based on the current state, receiving feedback in the form of a reward, and updating its policy accordingly. 
The goal of reinforcement learning (RL) is to learn a policy that maximizes rewards through iterative interactions with the environment \cite{DBLP:books/lib/SuttonB98}. 
The agent takes actions based on the current state, receives rewards, and updates its policy.
The trial-and-error learning nature of RL enables it to be a versatile method for making decisions in complex and dynamic environments \cite{vinyals2019grandmaster, lee2020learning, feng2023dense}.
RL algorithms can be broadly categorized into two types: value-based \cite{watkins1992q, sutton1995generalization, dabney2018distributional}        % mnih2015human, van2016deep, wang2016dueling,  
and policy-based \cite{williams1992simple, barto1983neuronlike, haarnoja2018soft}.      % mnih2016asynchronous, schulman2017proximal, lillicrap2015continuous, schulman2015trust,
Value-based methods estimate expected rewards for actions in a given state. 
Policy-based algorithms directly learn a policy model that maps states to actions and updates using Policy Gradient \cite{sutton1999policy}. 
Policy-based RL are popular for handling high-dimensional state and action spaces, and non-differentiable reward functions.
% Value-based methods learn a value function that estimates the expected reward of an action taken in a given state, while policy-based algorithms directly learn a policy model that maps states to actions and updates the policy using the Policy Gradient (PG) \cite{sutton1999policy}. Policy-based RL methods have gained traction in recent years due to their ability to handle high-dimensional state and action spaces, as well as non-differentiable reward functions.
Since the rewards in RL do not need to be differentiable with respect to model parameters, RL algorithms can be applied to a wide range of reward maximization problems \cite{DBLP:conf/cvpr/SiyaoYGLW0L022, ouyang2022training, pinto2023tuning}.        % zhou2018deep, jaques2016generating
Related to our work, \citet{sunco} propose a contrastive pre-trained reward to evaluate the correspondence between gesture and speech sequences and employs conservative Q-Learning (CQL) \cite{kumar2020conservative} for model optimization. 
Although state transitions and reward functions are obtainable, the method relies on offline RL, which limits the capability of the model.
Bailando \cite{DBLP:conf/cvpr/SiyaoYGLW0L022} use a hand-designed reward function to fine-tune the dance generation model. 
However, hand-designed reward functions require significant expert knowledge and make it difficult to comprehensively evaluate actions.
In our work, we fine-tune our diffusion model with online RL on the training set, using the learned reward model to refine the gesture generation model.

\section{Our Approach} %  4.5/4.75(5)-6.25 pages

\subsection{Multiple Skeletons Retargeting Network}

The structure of a skeleton is typically hierarchical \cite{DBLP:journals/tog/AbermanLLSCC20}, so we use graphs \cite{DBLP:journals/tvcg/WangHSZ21} to represent motions.
Connectivity is determined by the kinematic chains (the paths from the root joint to the end-effectors).
Nodes to represent corresponding joints and, in particular, leaf nodes to represent end-effectors.
The adjacency lists are expressed as $\mathcal{N}^d=\left\{\mathcal{N}_1^d, \mathcal{N}_2^d, \ldots, \mathcal{N}_J^d\right\}$, where $J$ is the number of joints and $\mathcal{N}_i^d$ denotes the edges whose distance in the tree is equal or less than $d$ from the $i$-th edge.

\subsubsection{Reference Pose Unification}       % Reference Poses and Root Processsing
The current full-body motion capture dataset for speech-driven gestures contains mainly: Trinity \cite{DBLP:conf/iva/FerstlM18} (244 min of audio, a male actor), ZEGGS \cite{DBLP:journals/cgf/GhorbaniFHTC23} (135 min of audio, a female actor, 19 different motion styles), BEAT \cite{DBLP:conf/eccv/LiuZIPLZBZ22} (76 hours, 30 speakers, 8 different emotions) and Talking with Hands \cite{DBLP:conf/iccv/LeeDMSSS19} (50 hours, two-person face-to-face conversations).
More details on the different skeletons can be found in the supplementary material.
Different gesture datasets have different reference poses and motion representations (number and position of joints). 
We take two datasets A and B as examples.
We first need to set the position and rotation of the unified reference poses, such as T-pose or A-pose.
We first centralize the reference representation to the root joints (e.g. the Talking with Hands dataset uses the `world' joint to maintain height).
Two reference poses $\mathbf{P}_A$ and $\mathbf{P}_B$ can be aligned through global and local translation and rotation:
\begin{equation}
\mathbf{P}_B=\mathbf{Q}^{A B} \mathbf{P}_A\left(\mathbf{Q}^{A B}\right)^{\top}
\end{equation}
where $\mathbf{Q}^{A B}$ denotes the reference poses transfer matrix.

The reference representation of a motion sequence of length $T$ based on reference pose $\mathbf{P}$ can be represented by 3D position and 4D rotation of the root joint as $\mathbf{R} \in \mathbb{R}^{T \times(3+4)}$.
The reference representations $\mathbf{R}$ of different skeletons can be retargeted after normalization according to the height of the reference pose.

\subsubsection{Motion Unification}
\begin{figure}[t]
  \centering
  \includegraphics[width=\linewidth]{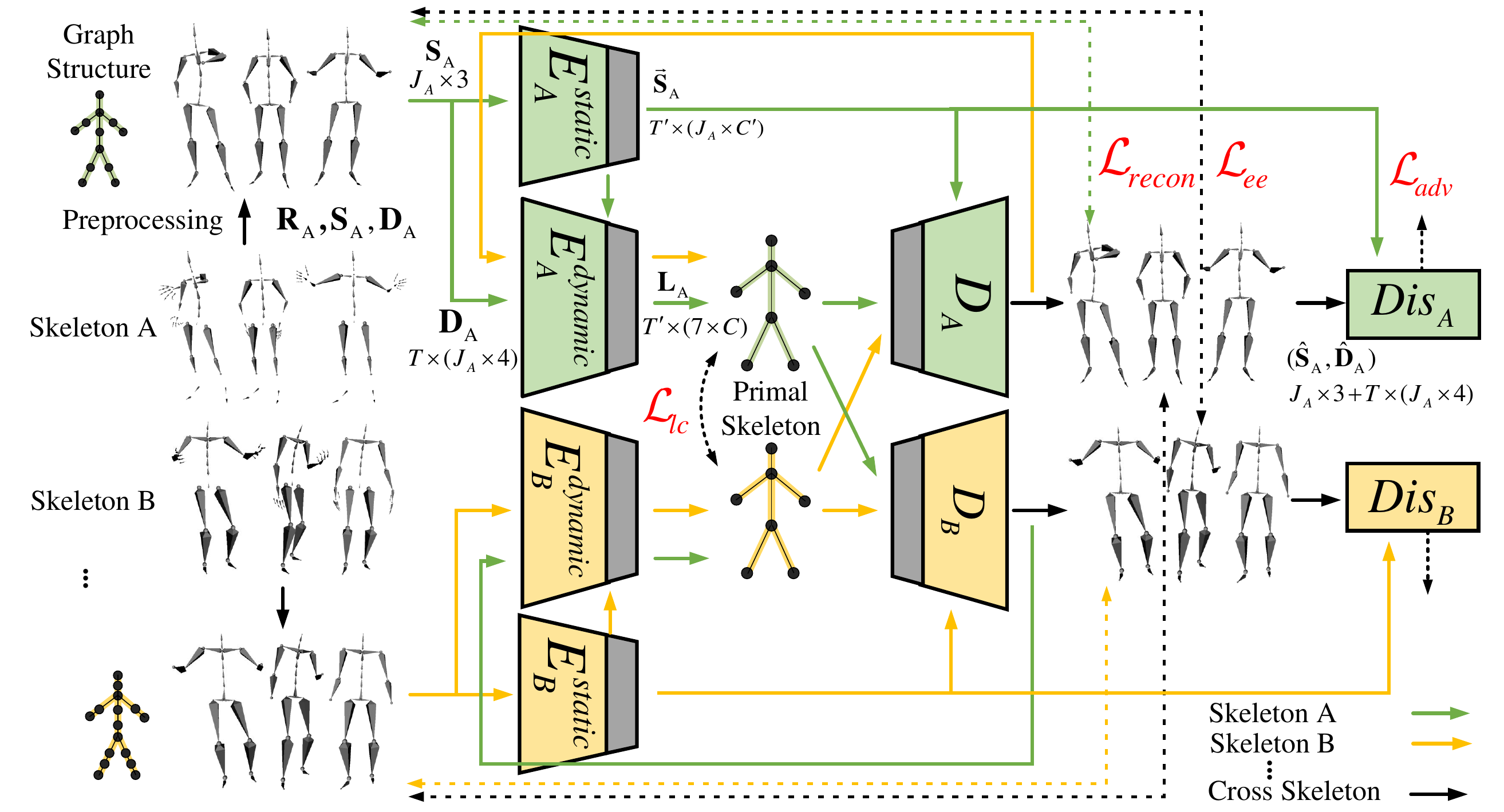}
  \caption{Motion is represented using a static encoder $E^{static}$ and a dynamic encoder $E^{dynamic}$. Assuming that all skeletons contain 5 end-effectors (2 hands, 2 feet, and head) and 2 mid-nodes, we use the latent representation $\mathbf{L}$ of the primal skeleton to represent all the different skeletons.}
  \Description{Motion is represented using a static encoder $E^{static}$ and a dynamic encoder $E^{dynamic}$. Assuming that all skeletons contain 5 end-effectors (2 hands, 2 feet, and head) and 2 mid-nodes, we use the latent representation $\mathbf{L}$ of the primal skeleton to represent all the different skeletons.}
    \label{retargeting}
\end{figure}
The motion of different skeletons consists of a static component $\mathbf{S} \in \mathbb{R}^{J \times3}$ (joint
offsets) and a dynamic one $\mathbf{D} \in \mathbb{R}^{T \times(J \times4)}$ (joint rotations).
To unify the motion of the different skeletons, we utilize a retargeting network architecture similar to \cite{DBLP:conf/aaai/ZhuYDW0L22}.
The architecture is shown in Figure \ref{retargeting}.
Here we take static component $\mathbf{S}_A$ and dynamic component $\mathbf{D}_A$ of skeleton A with $J_A$ joints as an example.
First, we adopt skeletal convolution and pooling layers \cite{DBLP:journals/tog/AbermanLLSCC20} in encoders to extract a deep latent representation $\mathbf{L_A}$ of the motion $\mathbf{S}$ and $\mathbf{D}$, which can be formulated as 
\begin{equation}
\overrightarrow{\mathbf{S_A}}={E^{static}_A(\operatorname{repeat}\left[\mathbf{S}_A\right]\times T^{\prime}})
\end{equation}
\begin{equation}
\mathbf{L_A} = E^{dynamic}_A(\mathbf{D}_A, \overrightarrow{\mathbf{S_A}})
\end{equation}
where operator `$\operatorname{repeat}$' denotes tiled and concatenated along the time dimension, $\overrightarrow{\mathbf{S_A}} \in \mathbb{R}^{T^{\prime} \times(J_A \times C')}$ and $\mathbf{L_A} \in \mathbb{R}^{T^{\prime} \times(7 \times C)}$, $T^{\prime}=T/d_{re}$, $d_{re}$ is the temporal down-sampling rate. 
Assuming that all skeletons contain 5 end-effectors (2 hands, 2 feet, and head) and 2 mid-nodes, we use the latent representation $\mathbf{L_A}$ of the primal skeleton to represent all the different skeletons, which contains only 7 nodes.
$C'$ and $C$ are the numbers of deep static and dynamic latent channels.

A following skeletal de-convolutional decoder $D_A$ projects $\overrightarrow{\mathbf{S_A}}$ and $\mathbf{L_A}$ back to the motion space as $\hat{\mathbf{S_A}}$, which can be formulated as
\begin{equation}
\label{retargeting_decoder}
(\hat{\mathbf{S}}_{\text{A}}, \hat{\mathbf{D}}_{\text{A}\rightarrow \text{A}}) = D_A(\mathbf{L_A}, \overrightarrow{\mathbf{S_A}})
\end{equation}
where $\hat{\mathbf{D}}_{\text{A}\rightarrow \text{A}}$ indicates that $\mathbf{L_A}$ is fed into the decoder $D_A$, simplified as $\hat{\mathbf{D}}_\text{A}$.

During training, $D_A$ tries to reconstruct the input motion, so the decoders are trained by minimizing the reconstruction losses:
\begin{equation}
\begin{aligned}
\mathcal{L}_{\text {rec }} & =\mathbb{E}\left[\left\|\hat{\mathbf{D}}_{\text{A}}-\mathbf{D}_\text{A}\right\|^2\right] +\mathbb{E}\left[\left\|\mathrm{FK}\left(\hat{\mathbf{D}}_{\text{A}}, \hat{\mathbf{S}}_A\right)-\mathrm{FK}(\mathbf{D}_\text{A}, {\mathbf{S}}_A)\right\|^2\right] 
\end{aligned}
\end{equation}
where operator `FK' is the forward kinematic to get the joint positions, which prevents the accumulaiton of error along the kinematic chain \cite{DBLP:journals/ijcv/PavlloFAG20}.

The skeletal-aware encodes enables retargeting motions of different skeletons into a common deep primal skeleton latent space. 
A latent consistency loss is applied to this shared representation to ensure that the retargeted motion retains the same dynamic features as the original clip:
\begin{equation}
\mathcal{L}_{\text {lc }}=\mathbb{E}\left[\left\|E^{dynamic}_{B}(\hat{\mathbf{D}}_{\text{A}\rightarrow \text{B}}, \overrightarrow{\mathbf{S_B}})-\mathbf{L_A}\right\|_1\right],
\end{equation}
where $\hat{\mathbf{D}}_{\text{A}\rightarrow \text{B}}$ indicates that $\mathbf{L_A}$ is fed into the decoder $D_B$.

Since different skeletons can share the same set of end-effectors (typically head, left hand, right hand, left foot, and right foot), the end-effectors of the original skeleton and the retargeted skeleton should have the same normalized velocity to avoid the artifact of re-targeting, such as foot sliding. 
This can be formulated as
\begin{equation}
\mathcal{L}_{\mathrm{ee}}=\mathbb{E}\sum_{e \in \mathcal{E}}\left\|\frac{V_{A_e}}{h_{A}}-\frac{V_{B_e}}{h_{B}}\right\|^2
\end{equation}
where $V_{A_e}$ and $V_{B_e}$ are the velocities of the $e$-th end-effector of skeletons A and B, respectively. 
$\mathcal{E}$ is the set of end-effectors.
$h_{A}$ and $h_{B}$ are the height of skeletons A and B, respectively.

And we use discriminator $Dis_{B}$ to evaluate whether the retargeted motion is plausible.
The adversarial loss can be formulated as
\begin{equation}
\mathcal{L}_{\mathrm{adv}}=\mathbb{E}\left[\left\|Dis_B\left(\hat{\mathbf{D}}_{A\rightarrow B}, \overrightarrow{\mathbf{S_B}}\right)\right\|^2\right]+\mathbb{E}\left[\left\|1-Dis_B\left(\hat{\mathbf{D}}_{\text{B}}, \overrightarrow{\mathbf{S_B}}\right)\right\|^2\right]
\end{equation}

The loss of the retargeting network can be computed as:
\begin{equation}
\label{loss_re}
\mathcal{L}_{re}=\mathcal{L}_{\text {rec }}+\lambda_{\text {lc }} \mathcal{L}_{\text {lc }}+\lambda_{\text {ee }} \mathcal{L}_{\text {ee }}+\lambda_{\text {adv }} \mathcal{L}_{\text {adv }}
\end{equation}

For details on the network structure, please refer to our supplementary material.

\subsection{Diffusion Model for Speech-driven Gesture Generation}
Diffusion models \cite{DBLP:conf/nips/HoJA20} have made great progress in motion generation \cite{DBLP:journals/corr/abs-2209-14916} due to their ability of to learn to gradually denoising starting from pure noise.
We unified the gestures by retargeting the skeletons of different gesture datasets to a primal skeleton, and now obtained a multi-deep primal skeleton gesture set $[\mathbf{L_A}, \mathbf{L_B}, ...]$ with the corresponding speech set $[\mathbf{A_A}, \mathbf{A_B}, ...]$.
To generate co-speech gestures with a diffusion model, we use DiffuseStyleGesture \cite{DBLP:conf/ijcai/myijcai}, which has recently achieved strong results on a single dataset, as our backbone model.
\begin{figure}[t]
  \centering
  \includegraphics[width=\linewidth]{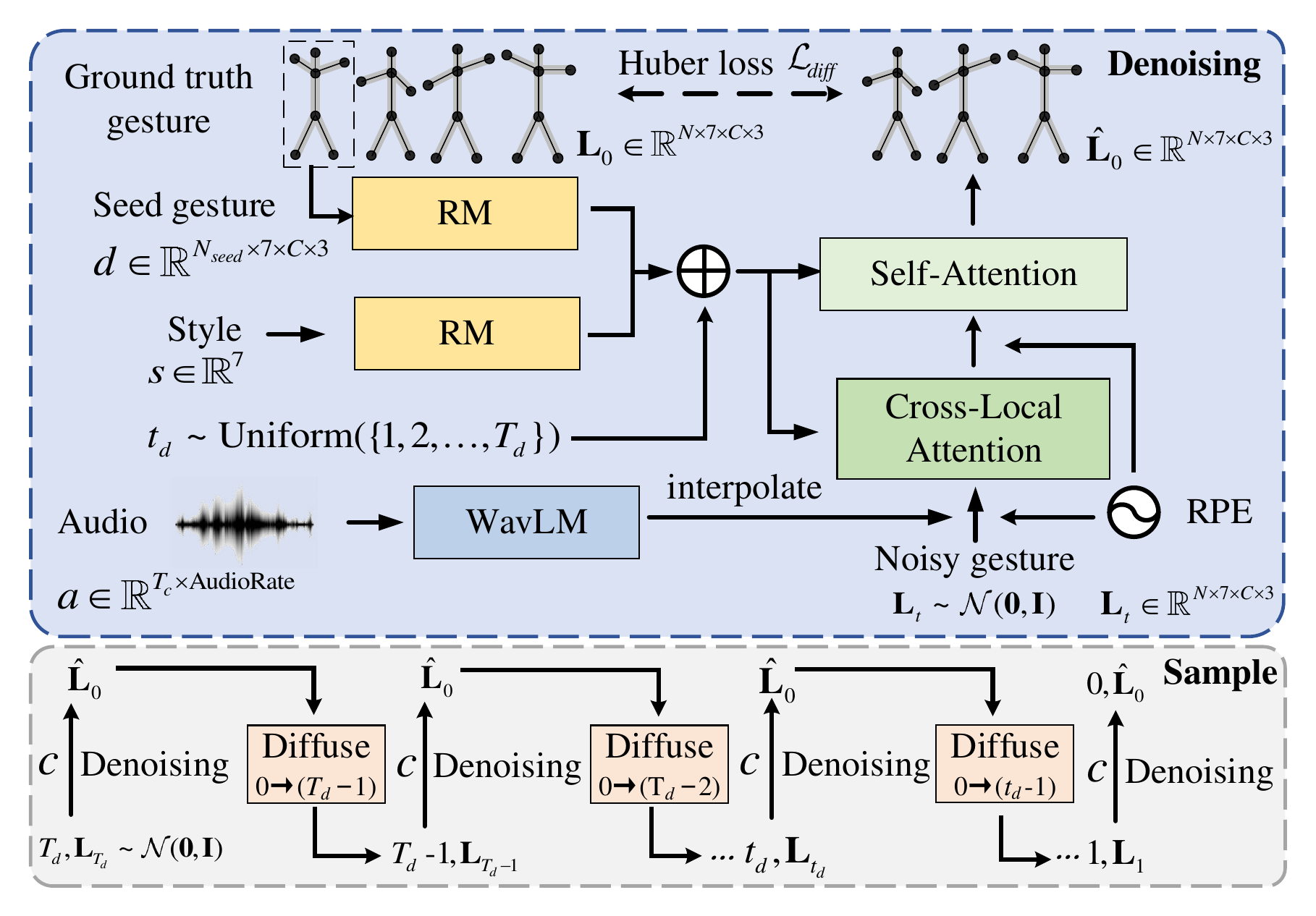}
  \caption{(Top) Denoising module.
   A noising step $t_d$ and a noisy gesture sequence $\mathbf{L}_t$ at this noising step conditioning on $c$ (including seed gesture $d$, style $s$ and audio $a$) are fed into the model. `RM' is short for random mask.
   (Bottom) Sample module.
   At each step $t_d$, we predict the $\hat{\mathbf{L}}_0$ with the denoising process based on the corresponding conditions, then add the noise to the noising step $\mathbf{L}_{t_d-1}$ with the diffuse process.
   This process is repeated from $t_d$ = $T_d$ until $t_d=0$.
   }
  \Description{(Top) Denoising module.
   A noising step $t_d$ and a noisy gesture sequence $\mathbf{L}_t$ at this noising step conditioning on $c$ (including seed gesture $d$, style $s$ and audio $a$) are fed into the model.
   (Bottom) Sample module.
   At each step $t_d$, we predict the $\hat{\mathbf{L}}_0$ with the denoising process based on the corresponding conditions, then add the noise to the noising step $\mathbf{L}_{t_d-1}$ with the diffuse process.
   This process is repeated from $t_D$ = $T_d$ until $t_d=0$.}
    \label{fig:diffusion}
\end{figure}
As shown in Figure \ref{fig:diffusion}, the diffusion model consists of two parts: the forward process (diffusion process) $q$ and the reverse process (denoising process) $p_\theta$.

We denote the generated gesture as $\mathbf{L}$ in the diffusion process, which has the same dimension as an observation data $\mathbf{L}_0 \sim q\left(\mathbf{L}_0\right)$, $q\left(\mathbf{L}_0\right)$ denotes the distribution of the real data $[\mathbf{L_A}, \mathbf{L_B}, ...]$.
According to a variance schedule $\beta_1, \beta_2, \ldots, \beta_{T_d}$ ($0<\beta_1<\beta_2<\cdots<\beta_{T_d}<1$, ${T_d}$ is the total time step),
we add Gaussian noise 
\begin{equation}
\label{Gaussian}
q\left(\mathbf{L}_{t_d} \mid \mathbf{L}_{{t_d}-1}\right)=\mathcal{N}\left(\mathbf{L}_{t_d} ; \sqrt{1-\beta_{t_d}} \mathbf{L}_{{t_d}-1}, \beta_{t_d} \mathbf{I}\right)
\end{equation}

In denoising process, the denoising process $p_\theta$ is a process of learning parameter $\theta$ via a neural network.
The noise $\mathbf{L}_{t_d}$ at time ${t_d}$ is used to learn $\mu_\theta$, $\Sigma_\theta$, then 
\begin{equation}
p_\theta\left({\mathbf{L}}_{{t_d}-1} \mid {\mathbf{L}}_{t_d}\right)=\mathcal{N}\left({\mathbf{L}}_{{t_d}-1} ; {\mu}_\theta\left({\mathbf{L}}_{t_d}, {t_d}\right), {\Sigma}_\theta\left({\mathbf{L}}_{t_d}, {t_d}\right)\right)
\end{equation}

\subsubsection{Denoising Module}
Our goal is to synthesize a gesture $\mathbf{L}^{1:N}$ of length $N$ given noising step $t_d$, noisy gesture $\mathbf{L}_{t_d}$ and conditions $c$ (including audio $a$, style $s$, and seed gesture $d$).     % arbitrary
% In our work, we follow \cite{ramesh2022hierarchical,tevet2022human} to predict the signal itself instead of predicting $\epsilon_\theta\left({\mathbf{L}}_{t_d}, {t_d}\right)$ \cite{ho2020denoising}.
\begin{equation}
\hat{\mathbf{L}}_0=\operatorname{Denoise}\left(\mathbf{L}_{t_d}, {t_d}, c\right)
\end{equation}

During training, noising step $t_d$ is sampled from a uniform distribution of $\{1, 2, \dots, T_{d}\}$, with the same position encoding as \cite{DBLP:conf/nips/VaswaniSPUJGKP17}.
Noisy gesture $\mathbf{L}_{t_d}$ has the same dimension as the real gesture $\mathbf{L}_0$ obtained by sampling from the standard normal distribution $\mathcal{N}(0,\mathbf{I})$. 
In the latent representation of the gesture we also extract the difference between two frames as latent velocity and also extract the difference between two frames of latent velocity as latent acceleration, therefore $\mathbf{L}_0 \in \mathbb{R}^{N \times(7 \times C \times 3)}$.
Audio features are generated from the pre-trained models of WavLM Large \cite{DBLP:journals/jstsp/ChenWCWLCLKYXWZ22}.
Then we use linear interpolation to align WavLM features and gesture $\mathbf{L}_0$ in the time dimension.
The styles of gestures are represented as one-hot vectors where only one element of a selected style is nonzero.
Seed gesture helps to make smooth transitions between consecutive syntheses \cite{DBLP:journals/tog/YoonCLJLKL20}.
The first $N_{seed}$ frames of the gestures clip are used as the seed gesture $d$ and the remaining $N$ frames are used as the real gesture $\mathbf{L}_0$ to calculate loss.
Self-attention \cite{DBLP:conf/nips/VaswaniSPUJGKP17} and cross-local attention \cite{DBLP:journals/tacl/RoySVG21} based on relative position encoding (RPE) \cite{DBLP:conf/iclr/KitaevKL20} are used to generate better speech-matched and realistic gesture.
Random masks (RM) are added to the pipeline of seed gesture $d$ and style $s$ feature processing for classifier-free learning \cite{DBLP:journals/corr/abs-2207-12598}.
During the training process, we combine the predictions of the conditional model $\operatorname{Denoise}\left(\mathbf{L}_{t_d}, t_d, c_1\right), c_1 = [d, s, a]$ and the unconditional model $\operatorname{Denoise}\left(\mathbf{L}_{t_d}, t_d, c_2\right), c_2 = [\varnothing, \varnothing, a]$:
% \begin{equation}
\begin{align}
\label{classifier-free}
\hat{\mathbf{L}}_{0{\gamma},{c_1},{c_2}}&=\gamma\operatorname{Denoise}\left(\mathbf{L}_{t_d}, t_d, c_1\right) + (1-\gamma)\operatorname{Denoise}\left(\mathbf{L}_{t_d}, t_d, c_2\right)
\end{align}
Then, as for style $s$ in condition, we can generate style-controlled gestures when sampling by interpolating or even extrapolating the two variants using $\gamma$, as $c_1 = [d, s_1, a], c_2 = [d, s_2, a]$ in Equation (\ref{classifier-free}).

The Denoising module can be trained by optimizing the Huber loss \cite{huber1992robust} between the generated poses $\hat{\mathbf{L}}_0$ and the ground truth human gestures $\mathbf{L}_0$ on the training examples:
\begin{equation}
\mathcal{L}_{diff}=\lambda_{diff}E_{\mathbf{L}_0 \sim q\left(\mathbf{L}_0 \mid c\right), {t_d} \sim[1, {T_d}]}\left[\operatorname{HuberLoss}(\mathbf{L}_0-\hat{\mathbf{L}}_0)\right]
\end{equation}      % \operatorname

\subsubsection{Sample Module}
The final co-speech gesture is given by splicing a number of clips of time duration $T_c$ with frame length $N$.
The initial noisy gesture $\mathbf{L}_{T_d}$ is sampled from the standard normal distribution and the other $\mathbf{L}_{t_d}({t_d}<{T_d})$ is the result of the previous noising step.
The seed gesture for the first clip can be generated by randomly sampling a gesture from the dataset or by setting it to the average gesture. 
Then the seed gesture for other clips is the last $N_{seed}$ frames of the gesture generated in the previous clip.
For every clip, in every noising step $t$, we predict the clean gesture $\hat{\mathbf{L}}_0$ =$\operatorname{Denoise}(\mathbf{L}_{t_d}, t_d, c)$, and add the noise to the noising step $\mathbf{L}_{t_d-1}$ using Equation (\ref{Gaussian}) with the diffuse process.
This process is repeated from $t_d$ = $T_d$ until $\mathbf{L}_0$ is reached (Figure \ref{fig:diffusion} bottom).
Please refer to our supplementary material for training details such as network structure and implementation details.

\subsection{Gesture Generation Refinement}

\subsubsection{Primal Gesture VQVAE}
Here we train a VQVAE to summarize meaningful gesture units to reduce the exploration space for following reinforcement learning.
Each code represents a unique gesture.
Besides, discrete spaces are more conducive to reinforcement learning for exploration \cite{DBLP:conf/aaai/Tang020, dulac2015deep}.
\begin{figure}[t]
  \centering
  \includegraphics[width=\linewidth]{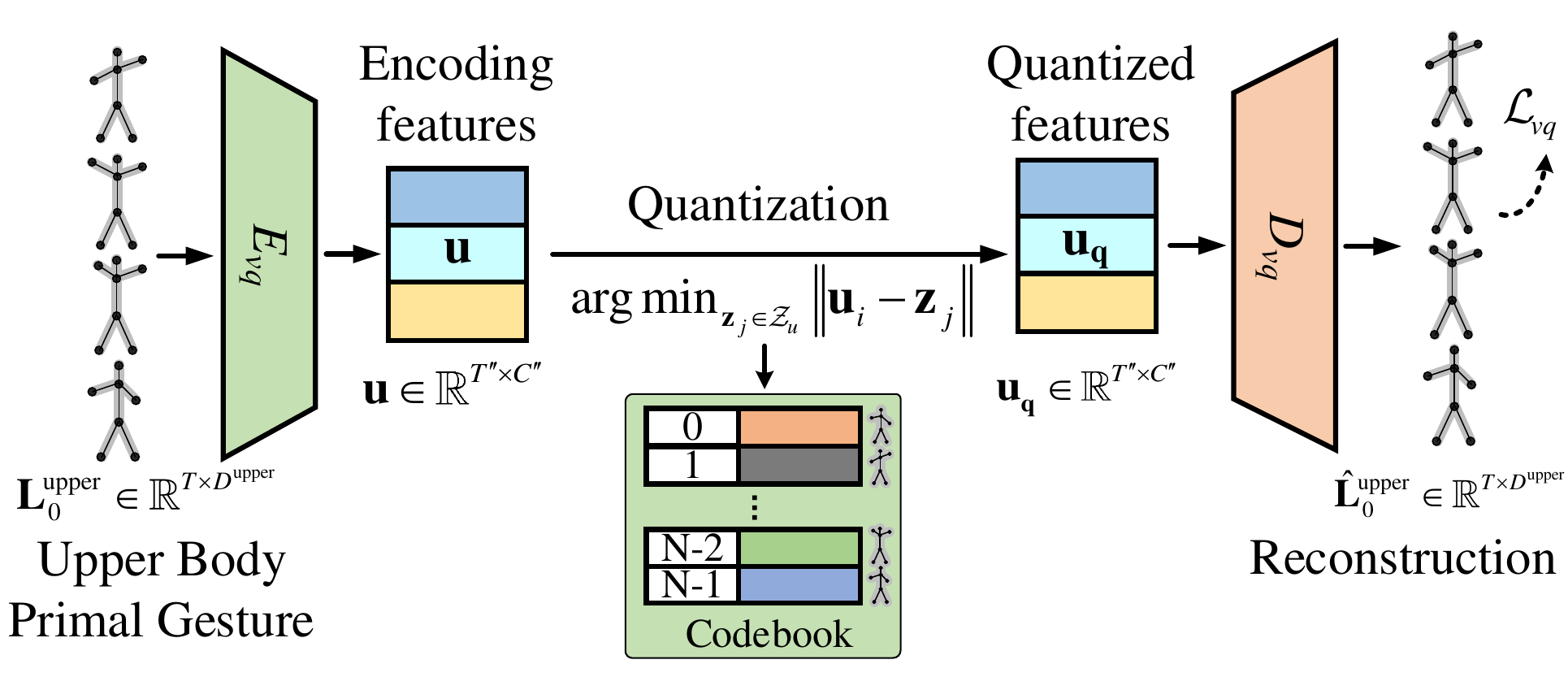}
  \caption{Structure of primal gesture VQVAE. After learning the discrete latent representation of the primal gesture of upper body, the gesture VQVAE encode and summarize meaningful gesture units.}
  \Description{Structure of primal gesture VQVAE. After learning the discrete latent representation of primal gesture of upper body, the gesture VQVAE encode and summarize meaningful gesture units, and reconstruct the target gesture sequence from quantized latent features.}
    \label{fig:vqvae}
\end{figure}
The architecture of the primal gesture VQVAE is shown in Figure \ref{fig:vqvae}.
Given the primal gesture sequence ${\mathbf{L}}_0^\text{upper}\in \mathbb{R}^{T \times D^\text{upper}}$ of the upper body, where $D^\text{upper}$ denotes primal gesture dimension of the upper body. 
We first adopt a 1D temporal convolution network $E_{vq}$ to encode the sequence $\mathbf{L}^\text{upper}_0$ to context-aware features $\mathbf{u}$
\begin{equation}
    \mathbf{u}=E_{vq}(\mathbf{\mathbf{L}}_0^\text{upper})
\end{equation}
where $\mathbf{u} \in \mathbb{R}^{T^{\prime\prime} \times C^{\prime\prime}}$ and $T^{\prime\prime}=T / d_{vq}$, $d_{vq}$ is the temporal down-sampling rate in VQVAE and $C^{\prime\prime}$ is the channel dimension of features.
Then we quantize $\mathbf{u}$ by mapping each temporal feature $\mathbf{u}_i$ to its closest codebook \cite{DBLP:conf/nips/OordVK17} element $z_j$ as $\mathbf{q}(.)$:
\begin{equation}
\mathbf{u}_{\mathbf{q}, i}=\mathbf{q}(\mathbf{u}) = \arg \min _{\mathbf{z}_j \in \mathcal{Z}_u}\left\|\mathbf{u}_i-\mathbf{z}_j\right\|
\end{equation}
where $\mathcal{Z}_u$ is a set of $C_b$ codes of dimension $n_z$. 
And $\mathbf{u}_{\mathbf{q}}$ is the elements of codebook $\mathcal{Z}_u$, $\mathbf{u}_q \in \mathcal{Z}_{u}$.
A following de-convolutional decoder $D_{vq}$ projects $\mathbf{u}_{\mathbf{q}}$ back to the deep latent space as a primal gesture sequence $\hat{\mathbf{L}}^\text{upper}_0$ for the upper body, which can be formulated as
\begin{equation}
\hat{\mathbf{L}}^\text{upper}_0=D_{vq}\left(\mathbf{u}_q\right)
\end{equation}

The VQVAE can be trained by optimizing $\mathcal{L}_{vq}$:
\begin{equation}
\label{VQVAE_loss}
\begin{split}
\mathcal{L}_{vq}&=\|{\hat{\mathbf{L}}^\text{upper}_0}-\mathbf{L}^\text{upper}_0\|_1+\alpha_1\left\|{\hat{\mathbf{L}}^{\text{upper}\prime}_0}-\mathbf{L}^{\text{upper} \prime}_0\right\|_1\\
&+\alpha_2\left\|{\hat{\mathbf{L}}^{\text{upper}\prime \prime}_0}-\mathbf{L}^{\text{upper}\prime \prime}_0\right\|_1+\left\|\operatorname{sg}[\mathbf{u}]-\mathbf{u}_{\mathbf{q}}\right\|+\beta_{vq}\left\|\mathbf{u}-\operatorname{sg}\left[\mathbf{u}_{\mathbf{q}}\right]\right\|
\end{split}
\end{equation}
where the first item is the reconstruction loss.
The next two items are velocity loss and acceleration loss \cite{DBLP:conf/cvpr/SiyaoYGLW0L022, DBLP:conf/cvpr/mycvpr}.
$\operatorname{sg}[\cdot]$ denotes the stop-gradient operation, and the term $\left\|\mathbf{u}-\operatorname{sg}\left[\mathbf{u}_{\mathbf{q}}\right]\right\|$ is the “commitment loss \cite{DBLP:conf/nips/OordVK17}” with weighting factor $\beta_{vq}$.

\begin{figure}[t]
  \centering
  \includegraphics[width=\linewidth]{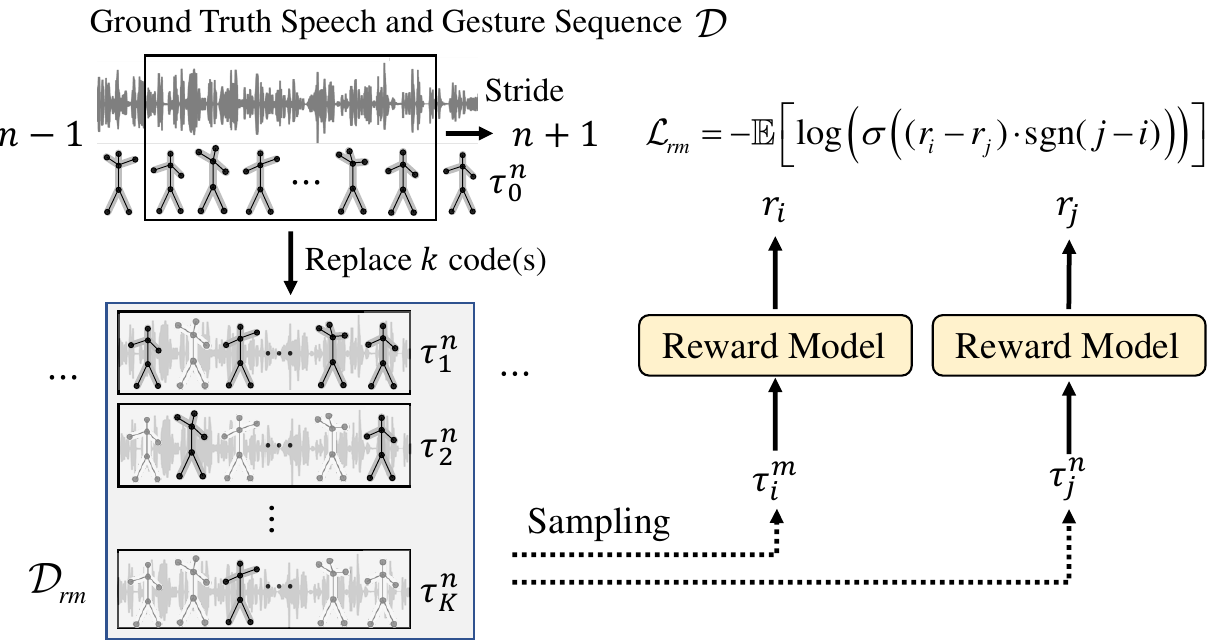}
  \caption{Reward model training. 
  We first sample a VQVAE-encoded speech-gesture pair, denoted as trajectory $\tau_0$.
  Then, we randomly replace $k$ gesture code(s) with random codes, where $k=1,\cdots,K$, resulting in $K$ speech-gesture trajectories with decreasing quality.
  Finally, we utilize the output of reward model $r$ to classify the trajectories with different qualities and optimize the reward model with the loss function $\mathcal{L}_{rm}$.
  }
  % \Description{rl.}
    \label{fig:rl}
\end{figure}
\subsubsection{Reinforcement Learning Finetuning}
To further enhance the alignment between the speech and gesture and increase the diversity of the generated gestures, we employed reinforcement learning to fine-tune the gesture generation model.
The reward signal is pivotal in balancing exploration and exploitation in reinforcement learning. Previous work \cite{DBLP:conf/cvpr/SiyaoYGLW0L022} attempts to optimize partial performance metrics of the model through hand-designed reward functions. 
However, in our experience, designing heuristic reward functions that comprehensively evaluate the model's performance is challenging. 
Reinforcement learning training is less stable than supervised learning, and if the reward function only considers specific metrics while neglecting others, the model's overall performance may deteriorate.

In this paper, we adopted Inverse Reinforcement Learning (IRL) \cite{ng2000algorithms} to learn a neural network model from human demonstrations to fit the true reward function and explain human behavior.
Specifically, our reward model training is shown in Figure \ref{fig:rl}, similar to \cite{brown2020better}. 
Firstly, we sample a speech-gesture pair from the VQVAE-encoded dataset $\mathcal{D}$, denoted as trajectory $\tau_0$. 
Then we randomly replace $k$ codes in the trajectory $\tau_0$ where $k=1,\cdots,K$ to generate $K$ trajectories $[\tau_1,\cdots,\tau_K]$. 
We sample $L$ tuples and thus get $L\times K$ trajectories to form the dataset $\mathcal{D}_{rm}$ to train the reward model.
We make a weak assumption that the more codes replaced with random codes, the worse the quality of the trajectories, including alignment with speech and diversity.
Then, we let the reward model $R_\psi$ classify these trajectories with different qualities (may come from different human demonstrations with different speech) $r = R_\psi(\tau)$ to determine which trajectory is better:
\begin{equation}
    \mathcal{L}_{rm}=-\mathbb{E}\left[\log \left(\sigma\left((r_i-r_j)\cdot \operatorname{sgn}(j-i)\right)\right)\right],
\end{equation}
where $\{i,j \in [1,\cdots,K], i \neq j\}$, $\sigma$ means the sigmoid function and
$\operatorname{sgn}$ means the signum function:
\begin{equation}
    \operatorname{sgn}(x)=
\begin{cases}
-1, & x<0 \\
1, & x>0
\end{cases}.
\end{equation}
By learning the classification task, the reward model can learn to output a scalar reward signal $r(\tau)=R_\psi(\tau)$ that makes reasonable evaluations on the quality of the trajectory $\tau$.

Given the reward model, we use the REINFORCE algorithm \cite{sutton1995generalization} to improve the model:

\begin{equation}
    \mathcal{L}_{RL}=-\mathbb{E}_{\tau\sim \pi}\left[\log p_{\pi}(\tau)r(\tau)\right],
\end{equation}
where $\pi$ means the current policy, \textit{i.e.}, the gesture model and $p_\pi(\tau)$ means the probability of $\tau$ given policy $\pi$.
During the fine-tuning process of the model, the reward model accurately scores the gesture under the given speech to improve alignment between speech and increase gesture diversity.

\subsubsection{Physics Guidance}
Inspired by \cite{DBLP:journals/corr/abs-2211-10658}, we consider that the foot should have contact with the ground when there is a left-right acceleration or an upward acceleration of the root.
% According to the foot contact with the ground, 
Then we use standard Inverse Kinematics (IK) optimization for physics guidance.
For more details please refer to the supplementary material.

\section{Experiments} % 5.5-7.75 pages

\subsection{Experiment Preparation}
\subsubsection{Implementation Details} 
We perform the training and evaluation on the Trinity \cite{DBLP:conf/iva/FerstlM18} and ZEGGS \cite{DBLP:journals/cgf/GhorbaniFHTC23} datasets.
Even based on motion capture, the hand quality is still low \cite{DBLP:journals/corr/abs-2301-05339, DBLP:journals/corr/abs-2211-09707, DBLP:conf/icmi/YoonWKVNTH22}, so we ignore hand motion currently.
Then the number of joints for the two datasets is $J_A=26$ and $J_B=27$, respectively.
We choose seven more typical and longest-duration styles (happy, sad, neutral, old, relaxed, angry, still) for training and validation.
For the Trinity dataset, there are no style labels and we consider all of its styles to be `neutral'.
And we divided the data into 8:1:1 by training, validation, and testing.
We first resample the motion of both datasets to 30fps.
All audio recordings are downsampled to 16kHz.
In terms of retargeting network, we set $d_{re}=4$, then the primal gesture is 7.5 fps.
We set all reference poses $\mathbf{R}$ to the T-pose at the origin with the foot in the Z-plane.
The dimension $C$ of each node of the primal gesture in latent space after convolution is 16.
We set $\lambda_{\text {lc }}=1$, $\lambda_{\text {ee }}=2$ and $\lambda_{\text {adv }}=0.25$ for Equation (\ref{loss_re}) and use the Adam \cite{DBLP:journals/corr/KingmaB14} optimizer with a batch size of 256 for 16000 epochs.
The retargeting network trained on an NVIDIA V100 GPU takes about 3 days.
While training the diffusion model and VQVAE, gesture data are cropped to a length of $N$ = 30 (4 seconds).
For the diffusion model, the Denoising module learns both the conditioned and the unconditioned distributions by randomly masking 10\% of the samples using Bernoulli masks.
The cross-local attention networks use 8 heads, 32 attention channels, 256 channels, the window size is 6, each window looks at the one window before it, and with a dropout of 0.1.
As for self-attention networks are composed of 8 layers, 8 heads, 32 attention channels, 256 channels, and with a dropout of 0.1.
We use the AdamW \cite{DBLP:conf/iclr/LoshchilovH19} optimizer (learning rate is 3$\times 10^{-5}$) with a batch size of 256 for 1000000 steps. 
Our models have been trained with $T$ = 1000 noising steps and a cosine noise schedule.
The diffusion model can be learned in about 3 days on one NVIDIA V100 GPU.
As for VQVAE, the size $C_b$ of codebook $\mathcal{Z}_u$ is set to 512 with dimension $n_z$ is 512.
We set the down-sampling rate $d_{vq}=2$.
And $\beta_{vq}$= 0.1, $\alpha_1$= 1 and $\alpha_2$= 1 for Equation (\ref{VQVAE_loss}).
we use the ADAM optimizer (learning rate is e-4, $\beta_1$ = 0.5, $\beta_2$ = 0.98) with a batch size of 128 for 200 epochs.
The VQVAE is learned on one NVIDIA A100 GPU for several hours. 
For more datasets and training details please refer to the supplementary material. 

\subsubsection{Evaluation Metrics}
Canonical correlation analysis (CCA) \cite{DBLP:journals/speech/SadoughiB19} is to project two sets of vectors into a joint subspace and then find a sequence of linear transformations of each set of variables that maximizes the relationship between the transformed variables. 
CCA values can be used to measure the similarity between the generated gestures and the real ones.
The closer the CCA is to 1, the better.
The Fréchet gesture distance (FGD) \cite{DBLP:journals/tog/YoonCLJLKL20} on feature space is proposed as a metric to quantify the quality of the generated gestures.
To compute the FGD, we trained an autoencoder to extract the feature.
Lower FGD is better.
Diversity \cite{DBLP:conf/cvpr/SiyaoYGLW0L022} in feature space is used to evaluate the diversity of the gestures.
We also report average jerk, average acceleration \cite{DBLP:conf/iva/KucherenkoHHKK19}, Hellinger distance \cite{DBLP:conf/icmi/KucherenkoJWHAL20}, and Beat Align Score \cite{DBLP:conf/cvpr/SiyaoYGLW0L022, DBLP:conf/cvpr/LiuWZXQLZWDZ22} in the supplementary material.

\subsection{Comparison to Existing Methods}
\subsubsection{Objective Evaluation}
\begin{table*}[!t]
\centering
\caption{Quantitative results on test set.
Bold indicates the best metric.
Among compared methods, StyleGestures \cite{DBLP:journals/cgf/AlexandersonHKB20}, Audio2Gestures \cite{DBLP:conf/iccv/0071KPZZ0B21}, ExampleGestures \cite{DBLP:journals/cgf/GhorbaniFHTC23}, and DiffuseStyleGesture \cite{DBLP:conf/ijcai/myijcai} are reproduced using officially released code with some optimized settings.
Objective evaluation is recomputed using the officially updated evaluation code \cite{DBLP:conf/iui/KucherenkoJYWH21, DBLP:conf/cvpr/SiyaoYGLW0L022}. Human-likeness and appropriateness are the results of MOS with 95\% confidence intervals.}
\label{tab:Quantitative}
\begin{tabular}{ccccccc}
\hline
\multirow{2}{*}{Name} & \multicolumn{4}{c}{Objective evaluation}                                     & \multicolumn{2}{c}{Subjective evaluation} \\ \cline{2-7} 
                      & Global CCA & CCA for each sequence & FGD $\downarrow$ & Diversity $\uparrow$ & Human-likeness      & Appropriateness     \\ \hline
Ground Truth        & 1.000          & 1.00 $\pm$ 0.00          & 0.0            & 10.03          & 4.22 $\pm$ 0.11 & 4.22 $\pm$ 0.11 \\
StyleGestures \cite{DBLP:journals/cgf/AlexandersonHKB20}       & 0.978          & \textbf{0.98 $\pm$ 0.01} & 15.89          & \textbf{13.86} & 3.56 $\pm$ 0.12 & 3.17 $\pm$ 0.13 \\
Audio2Gesture \cite{DBLP:conf/iccv/0071KPZZ0B21}      & 0.969          & 0.97 $\pm$ 0.01          & 19.78          & 6.148          & 3.61 $\pm$ 0.11 & 3.15 $\pm$ 0.14 \\
ExampleGestures \cite{DBLP:journals/cgf/GhorbaniFHTC23}     & 0.914          & \textbf{0.98 $\pm$ 0.01} & 10.49          & 5.418          & 3.77 $\pm$ 0.12 & 3.17 $\pm$ 0.14 \\
DiffuseStyleGesture \cite{DBLP:conf/ijcai/myijcai} & 0.987          & 0.97 $\pm$ 0.01          & 11.98          & 11.22          & 3.66 $\pm$ 0.12 & \textbf{3.46 $\pm$ 0.14} \\
Ours                & \textbf{0.988} & 0.95 $\pm$ 0.02          & \textbf{3.850} & 7.039          & \textbf{3.80 $\pm$ 0.11} & 3.42 $\pm$ 0.14 \\ \hline
\end{tabular}
\end{table*}

We compare our proposed model with StyleGestures \cite{DBLP:journals/cgf/AlexandersonHKB20}, Audio2Gestures \cite{DBLP:conf/iccv/0071KPZZ0B21}, ExampleGestures \cite{DBLP:journals/cgf/GhorbaniFHTC23}, and DiffuseStyleGesture \cite{DBLP:conf/ijcai/myijcai}.
The quantitative results are shown in Table \ref{tab:Quantitative}.
On the global CCA, our proposed model outperforms all other existing methods.
The highest global CCA shows a strong coupling between the generated gestures and the ground truth gestures.
CCA for each sequence is not as good as the other methods, and we suggest that this is because for each speech, the model learns the gestures across the skeleton.
Our method significantly surpasses the compared state-of-the-art methods with FGD, improves 6.64 (63\%) than the best compared baseline model ExampleGestures.
This shows the high quality of the generated gestures.
We can see that our model is not as good as StyleGesture in terms of Diversity.
The video results show that StyleGesture has a lot of cluttered movements, increasing diversity while decreasing human-likeness and appropriateness.
However, we would like to emphasize that objective evaluation is currently not particularly relevant for assessing gesture generation \cite{DBLP:conf/iui/KucherenkoJYWH21}. 
Subjective evaluation remains the gold standard for comparing gesture generation models \cite{DBLP:conf/iui/KucherenkoJYWH21, DBLP:journals/corr/abs-2303-08737}.
Current research on speech-driven gestures prefers to conduct only subjective evaluation \cite{DBLP:journals/corr/abs-2211-09707,DBLP:conf/ijcai/myijcai}.
Please refer to the supplementary video for more comparisons.

\subsubsection{User Study} 
To understand the real visual performance of our method, we conduct a user study among the gesture sequences generated by each compared method and the ground truth motion capture data. 
Following the evaluation in GENEA \cite{DBLP:conf/icmi/LuF22}, we evaluate human-likeness and gesture-speech appropriateness.
The length of the evaluated clips ranged from 22 to 50 seconds, with an average length of 35.4 seconds, as longer durations produce more pronounced and convincing appropriateness results \cite{DBLP:conf/icmi/Yang0LZLCB22}.
For human-likeness evaluation, each evaluation page asked participants “How human-like does the gesture motion appear?” 
In terms of appropriateness evaluation, each evaluation page asked participants “How appropriate are the gestures for the speech?” 
Participants rated at 1-point interval from 5 to 1, with labels (from best to worst) of ``excellent", ``good", ``fair", ``poor", and ``bad".
More details about the user study are shown in the supplementary material.
The mean opinion scores (MOS) on human-likeness and appropriateness are reported in the last two columns in Table \ref{tab:Quantitative}.

In terms of human-likeness, our model significantly surpasses the compared state-of-the-art methods.
However, it is not significantly different from ExampleGestures.
This is because ExampleGestures uses a reference gesture as `example' during inference, with already a priori knowledge of the gesture, sampling from a gesture distribution to get the generated gesture, so the human-likeness is strong.
For gesture and speech appropriateness, our model significantly outperforms StyleGestures, Audio2Gesture, and ExampleGestures, giving competitive results with DiffuseStyleGesture.
One reason for the gap compared to DiffuseStyleGesture is that DiffuseStyleGesture uses kinematic parameters such as the position, rotation angle, velocity, and rotation angular velocity of the root, as well as the position, rotation angle, velocity, rotation angular velocity, and gaze direction of each joint of the original motion as features of the gesture, which has a much larger dimension than the feature dimension of the primal skeleton gesture and may contain fine-grained skeletal details related to speech.
According to the feedback from the participants, our generated gestures are "more semantically relevant" and "more natural", while our method has "less power" compared to Ground Truth.
We suggest that this observation is due to the downsampling in the retargeting network and the VQVAE network. Smaller downsampling coefficients may result in faster and more powerful movements.
% However, this is consistent with current human subjective perception \cite{DBLP:conf/iui/KucherenkoJYWH21, DBLP:conf/icmi/YoonWKVNTH22} that speech-driven gestures lack proper objective metrics, even for FGD \cite{DBLP:journals/corr/abs-2212-04495,DBLP:journals/corr/abs-2303-08737}.

\subsection{Ablation Studies}
\subsubsection{Objective Evaluation}
\begin{table*}[!t]
\centering
\caption{Ablation studies results. '$-$' indicates modules that are not used. Bold indicates the best metric.}
\label{tab:Ablation}
\begin{tabular}{lcccccc}
\hline
\multicolumn{1}{c}{\multirow{2}{*}{Name}} & \multicolumn{4}{c}{Objective evaluation}                                     & \multicolumn{2}{c}{Subjective evaluation} \\ \cline{2-7} 
\multicolumn{1}{c}{}                      & Global CCA & CCA for each sequence & FGD $\downarrow$ & Diversity $\uparrow$ & Human-likeness      & Appropriateness     \\ \hline
Ground Truth & 1.000          & 1.00 $\pm$ 0.00          & 0.0            & 10.03     &     4.22 $\pm$ 0.11 & 4.22 $\pm$ 0.11 \\
Ours         & \textbf{0.988} & \textbf{0.95 $\pm$ 0.02} & 3.850 & \textbf{7.039} & {3.80 $\pm$ 0.11} & \textbf{3.42 $\pm$ 0.14} \\
\quad- RL         & 0.987          & 0.94 $\pm$ 0.03          & \textbf{3.132} & 7.008          & \textbf{3.82 $\pm$ 0.11} & 3.24 $\pm$ 0.16 \\
\quad- RL - VQVAE   & 0.987          & 0.94 $\pm$ 0.03          & 3.568          & 6.971          & 3.79 $\pm$ 0.11 & 3.33 $\pm$ 0.12 \\
\quad- Skeleton A & 0.972          & 0.94 $\pm$ 0.03 & 13.76         & 4.882     & 3.54 $\pm$ 0.12    & 3.00 $\pm$ 0.13  \\
\quad- Skeleton B & 0.965 & \textbf{0.95 $\pm$ 0.03} & 12.45         & 5.566 & 3.59 $\pm$ 0.13 & 3.09 $\pm$ 0.13 \\ \hline
\end{tabular}
\end{table*}

Moreover, we conduct ablation studies to address the performance effects of different components in the framework.
We performed the experiments on the following components: (1) reinforcement learning, (2) VQVAE, and (3) multiple skeletons.
The results of our ablation studies are summarized in Table \ref{tab:Ablation}.
% Supported by the results, when we do not use reinforcement learning, FGD got better, but other metrics got worse.
% This suggests that the quality of the generated gestures decreases after reinforcement learning.        % zerlin
The metrics on FGD indicate that after RL finetuning, the generated gestures have increased distance from the distribution of human gestures in the dataset, indicating that the model has explored some gestures that do not belong to the existing distribution of gestures in the dataset but are considered reasonable by the reward model. 
From the CCA and diversity metrics, it can be seen that the reward model can indeed generalize to gestures outside the dataset, allowing the model to generate more diverse and high-quality gesture movements that are not limited to the dataset.
When neither RL nor VQVAE is used, both FGD and diversity are still decreasing, which indicates the necessity of codebooks to generalize meaningful gestures.
When we use only a single dataset, we notice that both FGD and diversity decrease a lot, which indicates the essential importance of gesture generation for learning on multiple datasets.

\subsubsection{User Study}
Similarly, we conduct a user study of ablation studies.
The MOS on human-likeness and appropriateness are shown in the last two columns in Table \ref{tab:Ablation}.
In terms of human similarity, we can find that the scale of the dataset has a significant effect on the results, which demonstrates the importance of unifying the gesture dataset.
For speech and gesture appropriateness, it is also found that the scale of the dataset has the largest impact on this metric. 
Secondly, the appropriateness also decreased without reinforcement learning, shows the importance of data exploration.
The visual comparisons of this study can be also referred to the supplementary video.

\subsection{Diverse, Controllable, and Stylized Gesture Generation}
\begin{figure}[t]
  \centering
  \includegraphics[width=0.7125\linewidth]{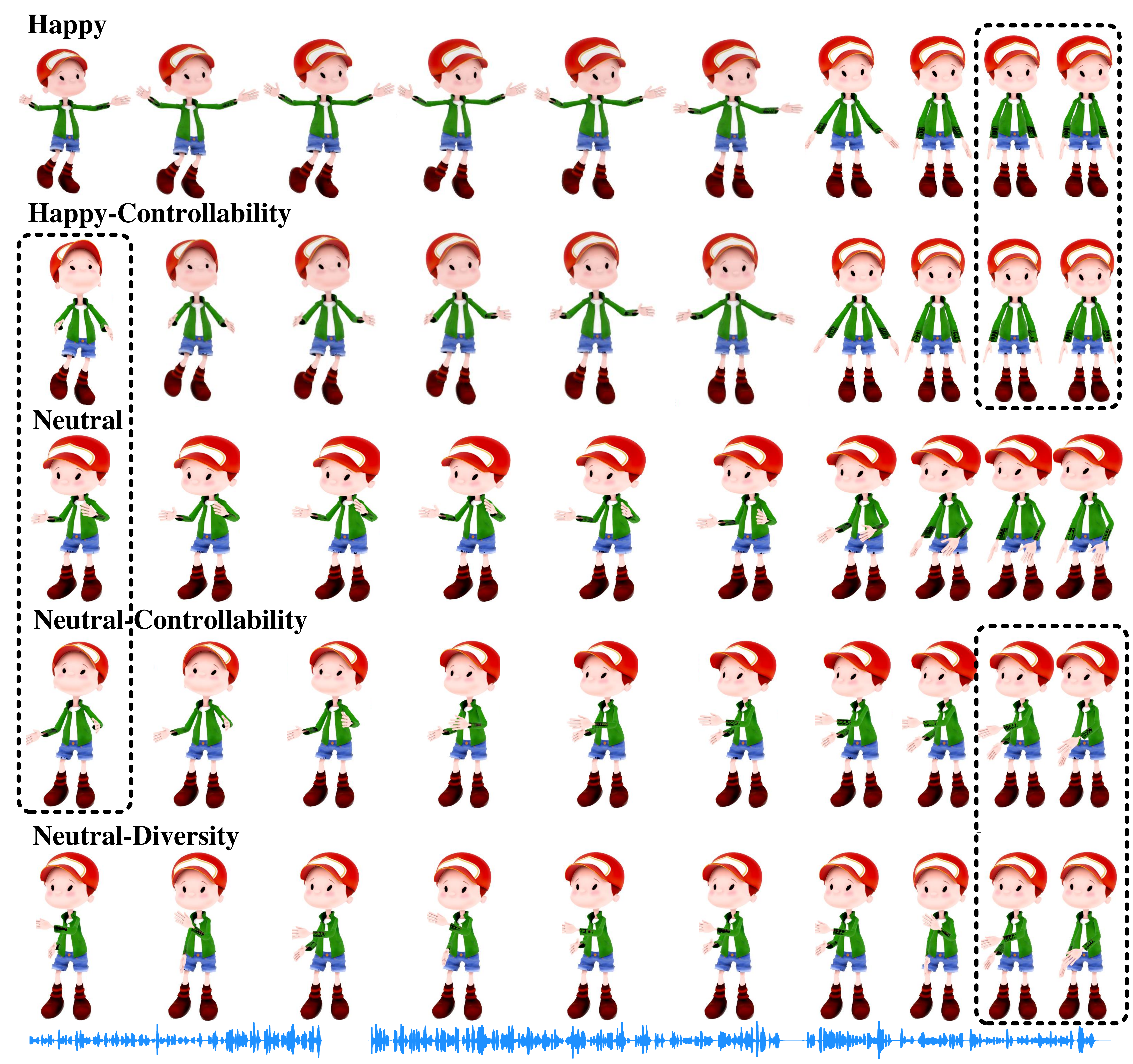}
  \caption{Visualization of the stylization, controllability, and diversity of generated gestures.
We randomly select a 2.67-second generated gesture clip (10 codes). 
Then setting $\gamma$ and $s$ in Equation (\ref{classifier-free}) to control the style and setting noisy gesture in diffusion model to generate diverse gestures.
The dashed boxes indicate that we control their code the same.
}
  \Description{Visualization of the stylization, controllability, and diversity of generated gestures.
We randomly select a 2.67-second generated gesture clip (10 codes). 
Then setting $\gamma$ and $s$ in Equation (\ref{classifier-free}) to control the style and setting noisy gesture in diffusion model to generate diverse gestures.
The dashed box indicates control their code the same.}
    \label{fig:exp-paper}
\end{figure}
\begin{itemize}
    \item \textbf{Stylization.}
        We can generate stylized gestures by setting $\gamma$ and $s$ in Equation (\ref{classifier-free}). 
        The intensity of the stylization can be controlled by the value of $\gamma$.
        As shown in Figure \ref{fig:exp-paper}, for the same speech, different styles of gestures can be generated while preserving matching with the speech.
    \item \textbf{Diversity.}
        Due to the diffusion model architecture, different noisy gesture and different seed gesture could generate different gestures even for the same speech and style, as shown in Figure \ref{fig:exp-paper}. 
        This is the same as real human speech, which creates diverse co-speech gestures related to the initial position.
    \item \textbf{Controllability.}
        Since we use VQVAE to generate gestures, it is easy to control the gesture or take out the code for interpretation.
        We can have a high level of control over speech-driven gestures at any time with the specified upper body code, as shown in the dashed box in \ref{fig:exp-paper}.
\end{itemize}
% \subsubsection{Stylization.}
%         We can generate stylized gestures by setting $\gamma$ and $s$ in Equation (\ref{classifier-free}). 
%         The intensity of the stylization can be controlled by the value of $\gamma$.
%         As shown in Figure \ref{fig:exp-paper}, for the same speech, different styles of gestures can be generated while preserving matching with the speech.
        
% \subsubsection{Diversity.}
%         Due to the diffusion model architecture, different noisy gesture and different seed gesture could generate different gestures even for the same speech and style, as shown in Figure \ref{fig:exp-paper}. 
%         This is the same as real human speech, which creates diverse co-speech gestures related to the initial position.
        
% \subsubsection{Controllability.}
%         Since we use VQVAE to generate gestures, it is easy to control the gesture or take out the code for interpretation.
%         We can have a high level of control over speech-driven gestures at any time with the specified upper body code, as shown in the dashed box in \ref{fig:exp-paper}.
For more details please refer to the supplementary material. 

\section{Discussion and Conclusion}

In this paper, we assume that the body gestures of the different skeletons
% weakly
% associated with speech 
are contained in the primal skeleton and present a unified gesture synthesis model for multiple skeletons.
UnifiedGesture demonstrates x major strength:
1) Benefit from using the skeleton-aware retargeting network to unify the different skeletons, while extending the dataset.
The model has stronger generalization.
And ablation experiments on a single skeleton effectively demonstrate that a larger amount of data can improve the performance of the model.
2) Based on a diffusion model, probabilistic mapping enhances diversity while enabling the generation of high-quality, speech-matched, and style-controlled gestures. 
3) VQVAE learns a codebook to summarize meaningful gesture units to improve controllability and interpretability. 
Reinforcement learning with a learned reward function helps refine the gesture generation model, enabling the model to explore the data and able to increase the diversity of the generated gestures.
The physics-based kinematic constraints also further improve gesture generation.
There is room for improvement in this research.
Besides speech, more modalities (e.g. text, facial expressions) could be taken into consideration to generate more appropriate gestures.
Solving the problem that the skeleton-aware encoder and decoder need to be re-trained for the new skeleton is also our future research direction.

\section*{Acknowledgments}

This work is supported by National Natural Science Foundation of China (62076144), Shenzhen Science and Technology Program (WDZC20220816140515001, JCYJ20220818101014030) and Shenzhen Key Laboratory of next generation interactive media innovative technology (ZDSYS20210623092001004).

% one skeleton train a encoder-decoder
% time-cos realtime

\bibliographystyle{ACM-Reference-Format}
\balance
% \bibliography{ACMMM2023}
\bibliography{ACMMM2023-longref}

\clearpage

\appendix

\section{Overview} % 1.5-2 pages    
   In this supplementary file, we present more experimental results analysis.
   \begin{itemize}
   \item We illustrate the differences between the skeletons of different gesture datasets.
   \item We give the specific structure of the retargeting network.
   \item We describe the implementation of reinforcement learning in detail.
   \item We give numerical results for more objective metrics.
   \item We show the design and scoring interface of the user study.
   \item We illustrate the model inference process.
   \end{itemize}
   
\section{Skeletons of Different Gesture Datasets}

\begin{figure}[!ht]
  \centering
  \includegraphics[width=0.7\linewidth]{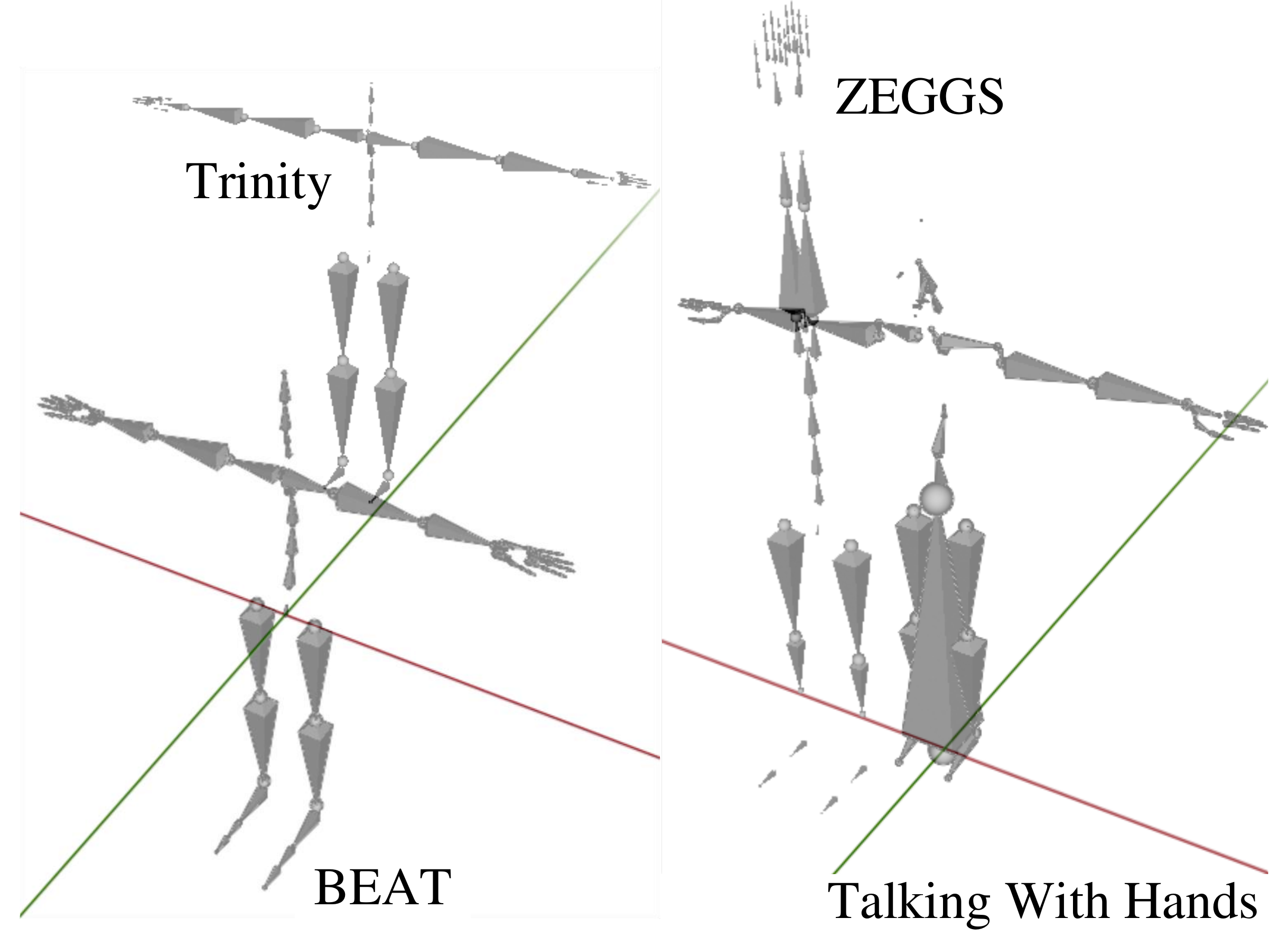}
  \caption{Visualization of different gesture datasets with reference poses, the position of their roots, and the number and location of their joints are different. The red and green lines represent the x-axis and y-axis, respectively.}
  \Description{Visualization of different gesture datasets with reference poses, the position of their roots, and the number and location of their joints are different. The red and green lines represent the x-axis and y-axis, respectively.}
    \label{different_motion_capture_standard}
\end{figure}

In practice, the dataset of the target skeleton is often small due to the expensive cost of motion capture, and we address the problem of how to utilize the existing datasets to improve the generation quality of the target skeleton. To this end, we design our algorithm to extend the dataset by unifying different skeletal datasets. We have explored the validity and necessity of unifying different skeletal datasets (Trinity and Zeggs), which is our main contribution, and our attempt takes an initial step towards generating gestures with big data and large motion models in the future.

Our current experimental setup is mainly to verify the feasibility of our idea, and we chose not to include larger datasets (e.g., Talking With Hands and BEAT) in the prior experiments due to the following two reasons: 1) As shown in Figure \ref{different_motion_capture_standard}, BEAT and Trinity have the same skeletal standard (Vicon), but Trinity, Zeggs, and Talking With Hands do not have the same skeleton. We experiment on Trinity and Zeggs, which are of comparable size, for the balance of the dataset distribution. 2) BEAT \cite{DBLP:conf/eccv/LiuZIPLZBZ22} has demonstrated better performance and generalization based on larger data compared to Trinity. That is, the more data with the same skeleton, the better the results. We solved the issue of adding data to improve performance from a different perspective, i.e., by adding more data with different skeletons. 

The finger mocap is still very challenging currently in the industry: 1) Optical motion capture systems, such as Vicon and OptiTrack, have occlusion problems, the markers on hand are easy to be occluded by hands; 2) Inertial motion capture systems, such as Xsens and Noitom, have the error accumulating problems, the inertial sensors are easy to accumulate errors; And 3) some deep learning-based pose estimation algorithms use optical motion capture or inertial motion capture as Ground Truth, which causes even lower accuracy.

We found the hand/finger quality of the existing mocap datasets \cite{DBLP:journals/corr/abs-2301-05339} is not good enough, especially when retargeted to an avatar. Datasets claimed with high-quality hand motion capture were still reported to have poor hand motion \cite{DBLP:journals/corr/abs-2301-05339}, e.g., ZEGGS Dataset in \cite{DBLP:journals/corr/abs-2211-09707} and Talking With Hands in \cite{DBLP:conf/icmi/YoonWKVNTH22}. So we ignore hand/finger motion currently, and leave it to future work.

The skeleton of the different gesture datasets is shown in Figure \ref{different_motion_capture_standard}.
These reference poses are, for the Trinity dataset, the root is not at the xy-plane origin; for the BEAT dataset, the root is at the origin, not the feet at the origin; for the ZEGGS dataset, the reference pose is not T-pose, in a straight line, and also not at the origin; for the Talking With Hands dataset, the skeleton structure is more complex, with many small skeletons and relying on a 'World' skeleton to maintain body height.
Like humans, in order to tell models what the references for retargeting are, we need to set up uniform reference gestures for them.

\section{Retargeting Network}

In this paper, we aim to solve the issue of co-speech generation, not retargeting. Ablation experiments of skeleton-aware networks (e.g. simple CNN) and comparisons with other baseline models are beyond the scope of our paper and have been done in \cite{DBLP:journals/tog/AbermanLLSCC20} to demonstrate the effectiveness of skeleton-aware networks for retargeting. Moreover, refinement after generating results is a general strategy, such as Bailando \cite{DBLP:conf/cvpr/SiyaoYGLW0L022}, GPT4 \cite{DBLP:journals/corr/abs-2303-08774}, and root and feet processing \cite{DBLP:journals/tog/AbermanLLSCC20, DBLP:conf/icmi/ZhouBC22}.

We consider the retargeting module, diffusion generation module, and refinement module in this work as an integrated pipeline, and we have demonstrated the effectiveness of each module in ablation experiments, and the refinement module based on reinforcement learning and physics guidance can help to achieve better performance.

\begin{figure}[!ht]
  \centering
  \includegraphics[width=0.83\linewidth]{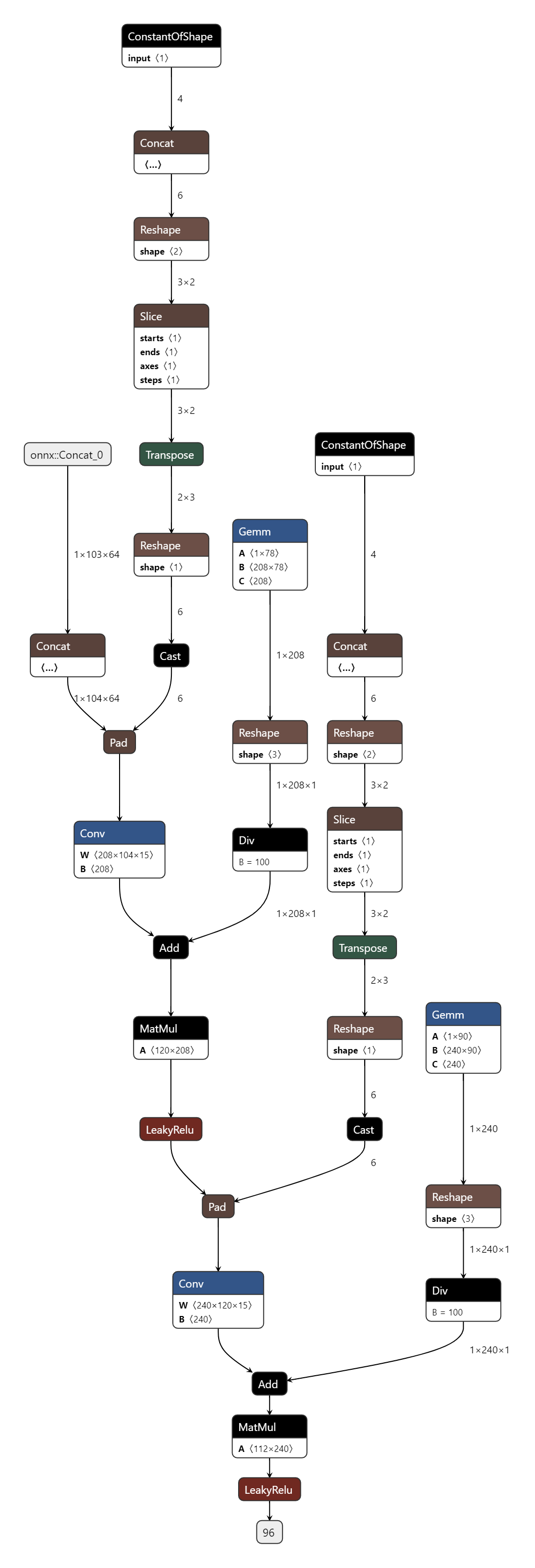}
  \caption{Visualization of the dynamic encoder $E^{dynamic}$ network structure. It mainly includes skeletal convolution, activation function, and skeletal pooling.}
  \Description{Visualization of the dynamic encoder $E^{dynamic}$ network structure. It mainly includes skeletal convolution, activation function, and skeletal pooling.}
    \label{encoder_onnx}
\end{figure}

The network structure of the dynamic encoder $E^{dynamic}$ is visualized in Figure \ref{encoder_onnx}.
It replicates the static offsets and concatenates them with dynamic motion, following skeletal convolution, LeakyRELU, and skeletal pooling.
The network structure of the static encoder $E^{static}$ and decoder $D$ is similar to it.
The decoder replaces the skeletal pooling with skeletal upsampling.

\section{Hyper-parameters of Reinforcement Learning}
\begin{table}[!ht]
\caption{Hyper-parameters of the reinforcement learning.}
\label{tab:hprm}
\begin{tabular}{c|c|c}
\toprule
                               & \textbf{Hyper-parameter} & \textbf{Value}        \\ 
\multirow{12}{*}{Reward Model} & Gesture Emb Dim          & 512                   \\ \hline
                               & Music Emb Dim            & 1024                  \\
                               & Hidden Dim               & 768                   \\
                               & Block Num                & 1                     \\
                               & Head Num                 & 12                    \\
                               & Context Length           & 18                    \\
                               & Causal                   & False                 \\
                               & Optimizer                & Adam                  \\
                               & Learning Rate            & 5e-4                  \\
                               & Weight Decay             & 2.5e-3                \\
                               & Batch Size               & 128                   \\
                               & K List                   & {[}0, 4, 9, 13, 18{]} \\ \hline
\multirow{4}{*}{REINFORCE}     & gamma                    & 1.0                   \\
                               & Max Grad Norm            & 0.1                   \\
                               & Optimizer                & AdamW                 \\
                               & Learning Rate            & 1e-6  
                               \\ \bottomrule
\end{tabular}
\end{table}
The hyper-parameter settings during our reinforcement learning process are shown in Table \ref{tab:hprm}.

\section{Objective Evaluation}

Regarding the models DisCo \citeA{liu2022disco} and Speech2AffectiveGestures \citeA{bhattacharya2021speech2affectivegestures}, only the upper body gestures were generated in their original papers. To generate gestures for the full body, these models need to learn kinematically relevant information, otherwise root sliding and drifting will occur. Text2gestures \citeA{bhattacharya2021text2gestures} is similar, its dataset is composed of the gestures of a seated, root-immobile human, and the ability to generate full-body free-motion gestures needs further validation. As for Mofusion \citeA{dabral2023mofusion} and GestureDiffuCLIP \citeA{ao2023gesturediffuclip}, they do not have open source code and are recently released work, so this work is not compared with them. The baseline models chosen in this paper are recently published and well behaved in full-body gestures.

\subsection{Comparison to Existing Methods}

In Table \ref{tab:Quantitative}, as for CCA, our model shows a slightly lower CCA for each sequence performance compared to some other methods, specifically StyleGestures and ExampleGestures. Our model achieves a CCA of 0.95, which is slightly lower than the 0.98 achieved by StyleGestures and ExampleGestures. The primary reason is that our model is designed to learn gestures across the entire skeleton, thereby emphasizing global coherence and alignment. This global approach may lead to compromises in capturing the detailed correlations at individual sequence levels. It is important to note that this trade-off is partly intentional since our aim was to excel in global CCA (0.988), which reflects the strong coupling between the generated gestures and the ground truth. With respect to diversity, our model scores 7.039, which is not as high as StyleGestures' 13.86. This is attributable to the fact that our model focuses on achieving human-likeness and appropriateness in gestures, which sometimes necessitates generating gestures that are more constrained and less varied. While StyleGestures focuses more on producing diverse gestures, it doesn't perform as well in the FGD (Frechet Gesture Distance) metric, indicating that its gestures may not be as high quality as those generated by our model.

The additional objective measures compared to the baseline model are shown in Table \ref{tab:supp_objective}.
Average Jerk, Average Acceleration, and Hellinger Distance are recomputed using \cite{DBLP:conf/iui/KucherenkoJYWH21}.
As for Beat Align Score, we use the method in \cite{DBLP:conf/iccv/0071KPZZ0B21} to calculate the beats of audio and follow \cite{DBLP:conf/cvpr/SiyaoYGLW0L022} to calculate the beats and diversity of gestures.
For Average Jerk and Average Acceleration, the closer to Ground Truth, the better.
For Hellinger Distance, the smaller the better.
Regarding Beat Align Score, the greater the better.
From the results, it can be seen that the gestures generated by our model are closest to the real velocity and acceleration distributions.
StyleGestures and DiffuseStyleGesture have motion velocity histogram distances that are more similar to the real gestures.
This could be caused by the more hand movements of both of them, please refer to our supplementary video.
DiffuseStyleGesture matches the beat of the speech better, which is consistent with the results of the human subjective evaluation.
Here we also want to emphasize that currently there is a lack of valid objective metrics for gesture generation and that subjective evaluation is the most effective \cite{DBLP:conf/iui/KucherenkoJYWH21, DBLP:journals/corr/abs-2303-08737}. Please refer to our video for further visualization and comparison.

\begin{table*}[!ht]
\centering
\caption{Additional quantitative results on the test set. Bold indicates the best metric, e.g., closest to Ground Truth or minimal, etc.}
\label{tab:supp_objective}
\begin{tabular}{ccccc}
\hline
Name                & Average Jerk             & Average Acceleration   & Hellinger Distance $\downarrow$ & Beat Align Score $\uparrow$ \\ \hline
Ground Truth        & 745.42 $\pm$ 661.80         & 64.50 $\pm$ 46.93         & 0.0                & 0.191            \\
StyleGestures \cite{DBLP:journals/cgf/AlexandersonHKB20}       & 7667.90 $\pm$ 3609.80       & 319.40 $\pm$ 96.62        & \textbf{0.139}     & 0.156            \\
Audio2Gesture \cite{DBLP:conf/iccv/0071KPZZ0B21}      & 8460.02 $\pm$ 578.60        & 306.10 $\pm$ 25.91        & 0.151              & 0.123            \\
ExampleGestures \cite{DBLP:journals/cgf/GhorbaniFHTC23}     & 90.82 $\pm$ 5.69            & 18.81 $\pm$ 0.76          & 0.149              & 0.157            \\
DiffuseStyleGesture \cite{DBLP:journals/tog/YoonCLJLKL20} & 1958.61 $\pm$ 316.69        & 204.71 $\pm$ 29.64        & \textbf{0.139}     & \textbf{0.239}   \\
Ours                & \textbf{263.87 $\pm$ 70.95} & \textbf{28.66 $\pm$ 9.26} & 0.141              & 0.166            \\ \hline
\end{tabular}
\end{table*}

\subsection{Ablation Studies}

In Table \ref{tab:Ablation}, we observe that when the RL (Reinforcement Learning) component is removed from our model (Ours - RL), the FGD (Frechet Gesture Distance) decreases from 3.850 to 3.132. This indicates that the gestures generated without RL are closer to the distribution of human gestures in the dataset. However, the slightly better FGD score does not necessarily represent better generalization. RL is essential for enabling the model to explore beyond the dataset and generate gestures that, though slightly further from the human distribution, are more diverse and considered reasonable by the reward model, as can be seen from the Global CCA and Diversity metrics. Therefore, while the ablated version without RL shows better FGD, the trade-off is in generalization and diversity. When neither RL nor VQVAE (Vector Quantized Variational AutoEncoder) is used (Ours - RL - VQVAE), the FGD is higher than Ours - RL, but still lower than our full model. The absence of VQVAE causes a reduction in diversity. This suggests that the VQVAE module helps to generate meaningful gestures. In this ablation, without the RL module, the model is not encouraged to explore beyond the dataset, and without VQVAE, it struggles to generalize meaningful gestures. The combination of RL and VQVAE in the full model ensures that meaningful gestures are generated, and the model is encouraged to explore beyond the dataset, which enhances the diversity and quality of the generated gestures. The ablation studies with - Skeleton A and - Skeleton B demonstrate the importance of having diverse training datasets. As we can see, removing either Skeleton A or Skeleton B increases the FGD dramatically to 13.76 and 12.45 respectively. This indicates that the model is not generalizing well without diverse training data. Similarly, Diversity is significantly reduced when training on a single dataset, highlighting the importance of training on multiple datasets for producing varied and high-quality gestures. Our full model, incorporating RL, VQVAE, and multiple datasets, achieves a balance across these aspects, as is evident in the objective and subjective evaluations.

Similarly, we calculated more objective measures of the ablation experiment.
The results are shown in Table \ref{tab:supp_objective_ab}.
As can be seen from the results, all four metrics even get better when the model does not use reinforcement learning as well as VQVAE; when the model removes reinforcement learning and VQVAE, the average velocity, average acceleration, and velocity distribution histogram achieves the best case.
When the model was trained on a single dataset skeleton only, the average velocity and average acceleration decreased more significantly, but the other two metrics showed the opposite trend.
These are contradictory to the subjective evaluation and other objective evaluation metrics, so we argue with the results of these objective metrics, but we still report these results.
We believe that subjective evaluation is still the most convincing method for now, and that objective metrics are all still inconsistent with subjective human perception. Please refer to the video for further comparison.

\begin{table*}[!ht]
\centering
\caption{Additional ablation studies results. '$-$' indicates modules that are not used. Bold indicates the best metric.}
\label{tab:supp_objective_ab}
\begin{tabular}{lcccc}
\hline
\multicolumn{1}{c}{Name} & Average Jerk             & Average Acceleration   & Hellinger Distance $\downarrow$ & Beat Align Score $\uparrow$ \\ \hline
Ground Truth & 745.42 $\pm$ 661.80 & 64.50 $\pm$ 46.93 & 0.0   & 0.191          \\
Ours         & 263.87 $\pm$ 70.95  & 28.66 $\pm$ 9.26  & 0.141 & 0.166          \\
\quad- RL         & 270.63 $\pm$ 75.24  & 29.30 $\pm$ 9.56  & 0.135 & 0.173          \\
\quad- RL - VQVAE             & \textbf{276.83 $\pm$ 76.98} & \textbf{29.62 $\pm$ 9.69} & \textbf{0.126}     & 0.172            \\
\quad- Skeleton A & 259.49 $\pm$ 67.10  & 26.20 $\pm$ 8.69  & 0.133 & \textbf{0.176} \\
\quad- Skeleton B & 232.15 $\pm$ 52.80  & 20.45 $\pm$ 6.64  & 0.155 & 0.162          \\ \hline
\end{tabular}
\end{table*}

\section{User Study}

Human-likeliness and Appropriateness for gesture scoring are the two dimensions that have been used in the gesture generation (GENEA) Challenge \cite{DBLP:conf/iui/KucherenkoJYWH21, DBLP:journals/corr/abs-2303-08737, DBLP:conf/icmi/YoonWKVNTH22} and are currently the dominant metrics in gesture generation. Some work in user study has different focus on different topics, such as user evaluation diversity \cite{DBLP:conf/iccv/0071KPZZ0B21}, consistency \citeA{zhuang2023gtn}, stylization \cite{DBLP:conf/ijcai/myijcai}, appropriateness of interaction with the listener added in this year's gesture generation Challenge \citeA{kucherenko2023genea}, etc. However, Human-likeliness and Appropriateness are dimensions that are used in almost all user studies of gesture generation methods, so we followed these two metrics. To analyze user studies using statistics, we used MOS with 95\% confidence intervals to represent the results of each metric. If there is no overlap in the 95\% confidence intervals of the ratings between the different models, then the difference is considered to be statistically significant.

\begin{figure}[!ht]
  \centering
  \includegraphics[width=\linewidth]{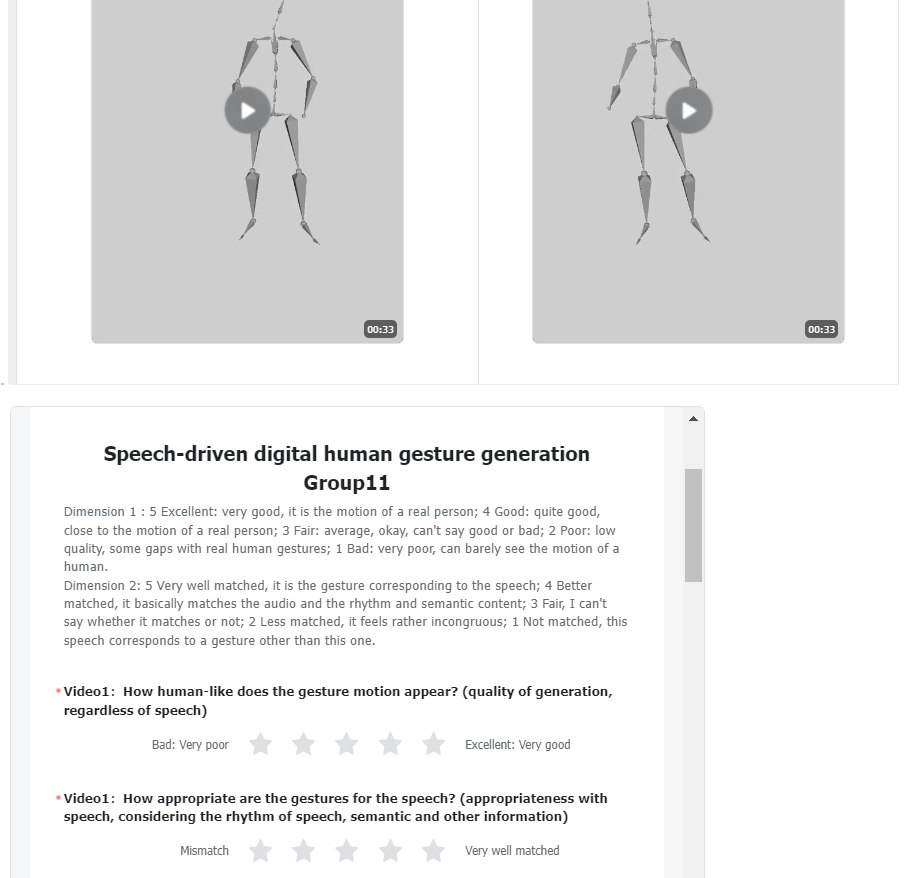}
  \caption{A screenshot of the subjective evaluation scoring interface.}
  \Description{A screenshot of the subjective evaluation scoring interface.}
    \label{shot}
\end{figure}

We put all the generated primal skeleton gestures through the decoder of the ZEGGS dataset skeleton to generate the final gestures.
The generated motion capture file (bvh) is rendered by Blender \cite{Blender} and the camera stays still.
Before starting the evaluation, we told participants, for each Video, two dimensions were evaluated:
1. naturalness (human-likeness), i.e., the quality of the generated motion, without considering speech.
2. suitability (appropriateness), i.e., the relationship between the generated gestures and the speech, considering the speech, e.g., whether the gestures match the audio rhythm, or the text semantics.
We asked people to ignore the influence of the hands on the scoring and to focus only on the skeletal movements of the whole body.
A total of 31 individuals took part in the subjective evaluation scoring, with 2 subjects between the ages of 40 and 50 and the rest between the ages of 20 and 30. 
About 85\% of the participants were male and 15\% were female. They were all good English speakers.
For both the comparison with the baseline model and the ablation experiments, we selected 10 segments of audio to be scored with their generated gestures. 
Five segments are male voices from the Trinity dataset, and another 5 segments are female voices from the ZEGGS dataset. 
Ten models in total were scored.
For each model, there were 10 segments (approximately 5 minutes in total) of audio with generated gestures. 
We paid each participant an hourly rate of approximately 10 USD, which is above the average salary level \cite{DBLP:journals/corr/abs-2303-08737}.
A screenshot of the subjective evaluation scoring screen is shown in Figure \ref{shot}.

\section{Gesture Generation for Multiple Skeletons}

Take the general gesture generation of skeleton A as an example.
 $\hat{\mathbf{L}}^\text{upper}_0$ and ${\mathbf{L}}^\text{lower}_0$ are the upper body of primal gesture sequence after VQVAE reconstruction and the lower body of the diffusion model output, respectively.
The generated gestures are finally given by Equation (4) as $D_A[(\hat{\mathbf{L}}^\text{upper}_0, {\mathbf{L}}^\text{lower}_0), \overrightarrow{\mathbf{S_A}}]$.
Similarly, the primal gesture sequence is fed into the decoder of whichever skeleton the gesture is generated, as shown in Figure \ref{framework}.

\section{More Discussion}

\begin{figure}[!ht]
  \centering
  \includegraphics[width=\linewidth]{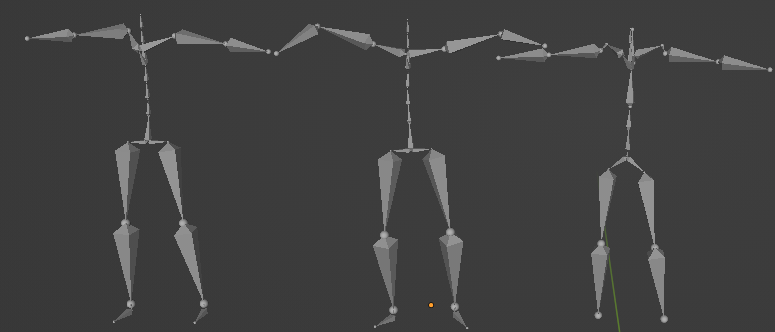}
  \caption{Retargeting visualization on BEAT and TWH. On the left is the result using Auto-rig in Blender, in the middle is the real motion, and on the right is the result generated by the retargeting network.}
  \Description{Retargeting visualization on BEAT and TWH. On the left is the result using Auto-rig in Blender, in the middle is the real motion, and on the right is the result generated by the retargeting network.}
    \label{shot-TWH}
\end{figure}

We obtained similar results from our experiments on BEAT and TWH. The criteria for the motions in the BEAT and Trinity datasets are the same, and TWH is slightly more complex, especially for the shoulder joints, as shown on the right side of the Figure \ref{shot-TWH}. From the figure, we can find that the three poses are generally similar, because the retargeting network is constrained with terminal positions and uses 5 terminals + 2 intermediate joints for a total of 7 joints as the middle representation, so more detailed information may be neglected. For example, the details of elbow and shoulder.

We appreciate the concerns regarding the scalability of our proposed method as new skeleton data is incorporated, necessitating the retraining of the network. We agree that it is theoretically feasible to decouple the problem into two separate parts - one focusing only on learning a uniform skeleton representation for the autoencoder system, and the other focusing only on co-speech gesture synthesis. This approach allows only the retargeting part of the network to be retrained when new skeleton data is added, resulting in significant computational cost and time savings. However, it is also important to acknowledge that while these two problems can be technically decoupled, there may be complex interactions between them in practice. There are indeed a lot of works \cite{DBLP:journals/jvca/KimJK20, DBLP:journals/corr/abs-2212-02837, DBLP:conf/iva/KucherenkoHHKK19, DBLP:conf/cvpr/VillegasYCL18} on learning to retarget between different skeletons; or on learning co-speech gesture generation \cite{DBLP:conf/iva/AlexandersonSHK20, DBLP:conf/cvpr/GinosarBKCOM19, DBLP:conf/cvpr/LiuWZXQLZWDZ22, DBLP:conf/icmi/LuF22}. We are the first approach to attempt to integrate the both, and the integration of retargeting with speech-driven gestures can yield impressive results.

\bibliographystyleA{ACM-Reference-Format}
% \balance
% \bibliography{ACMMM2023}
\bibliographyA{ACMMM2023-longref}

\end{document}